\documentclass[11pt]{article}

\usepackage{amsmath}
\usepackage{caption}
\usepackage{subcaption}
\usepackage[margin=2.5cm]{geometry}
\usepackage{bm}
\usepackage{amsfonts}
\usepackage{graphicx}
\usepackage{hyperref}
\usepackage{amssymb}

\hypersetup{
    colorlinks=true,
    linkcolor=magenta,
    filecolor=magenta,
    urlcolor=magenta,
}

\title{Analysis of tensor-product discontinous Galerkin operators for Vlasov-Poisson simulations and GPU implementation on Python}
\author{D.W. Crews}
\date{\today}

\begin{document}
\maketitle

\begin{abstract}
    The discontinuous Galerkin (DG) finite element method is conservative, lends itself well to parallelization,
    and is high-order accurate due to its close affinity with the theory of quadrature and orthogonal polynomials.
    When applied with an orthogonal discretization (\textit{i.e.} a rectilinear grid) the DG
    method may be efficiently implemented on a GPU in just a few lines of high-level language such as Python.
    This work demonstrates such an implementation by writing the DG semi-discrete equation in a tensor-product form
    and then computing the products using open source GPU libraries.
    The results are illustrated by simulating a problem in plasma physics, namely an instability in the magnetized
    Vlasov-Poisson system.
    Further, as DG is closely related to spectral methods through its orthogonal basis it is possible to calculate
    a transformation to an alternative set of global eigenfunctions for purposes of analysis or to perform additional
    operations.
    This transformation is also posed as a tensor product and may be GPU-accelerated.
    In this work a Fourier series is computed for example (although this does not beat discrete Fourier transform), 
	and is used to solve the Poisson part of the Vlasov-Poisson system to $\mathcal{O}(\Delta x^{n+1/2})$-accuracy.
\end{abstract}

\section{Introduction}\label{sec:intro}
In science and engineering a typical problem is to model the change of some quantities in time, such as pressure
and density in a fluid flow or the distribution of temperature in a heated object.
When the underlying physics result in linear equations then solutions can be found.
Yet in applications these models usually take the form of nonlinear partial differential equations (PDEs).
To name a few there are
the reaction-diffusion systems of chemistry; the famous Navier-Stokes equations; 
the Hamilton-Jacobi equation of mechanics and optimal control; reduced wave equations of optics, seismology, and oceanography;
field equations like Maxwell's and Einstein's; Boltzmann's kinetic equation; 
quantum theory's Schr\"{o}dinger equation; etc. So it is useful to solve nonlinear PDEs.

Lacking a well-understood technique to solve nonlinear equations, PDEs are often simulated numerically by
representing the solution on some set of points.
These systems evolve in many dimensions $d$, typically of $d\geq 3$,
so in solving them one experiences the ``curse of dimensionality."
This means that if $N$ points are used to represent the solution in any one direction then the total number of points is $N^d$.
To remain computationally tractable one is limited to a certain total $M$, so the approximately $M^{1/d}$ points of each
dimension should be chosen judiciously to obtain the most accurate solution, provided by \textit{e.g.} quadrature nodes.

The discontinuous Galerkin (DG) finite element method combines a prudent choice of points with a highly
parallelizable approach~\cite{hesthaven}.
An excellent review of DG in a historical context is given in~\cite{cockburn2}.
Many expositions are precisely formulated and, rightly so, are heavy with mathematical
formalisms such as function spaces and various error estimates~\cite{dolejsi}~\cite{johnson}.
Additionally, the DG literature pushes the boundaries of the method with hybridizable~\cite{rhebergen},
semi-Lagrangian~\cite{dongmi}, superconvergent~\cite{castillo}, and space-time~\cite{tavelli} innovations.
Studies show impressive performance and scaling of DG-type methods on GPUs~\cite{einkemmer}.
Yet there seems to be a gap in the recent literature, namely an easy-to-understand description of a vanilla DG method
on a GPU.
This article intends to fill this gap with a cogent explanation of an implementation easily performed
on a desktop with a graphics card.
Ideally this will complement the existing literature for beginners with the method. 

The DG method is based on a discontinuous piecewise interpolation through points in a collection of elements.
The PDE is written in an integral (weak) form and then projected onto the polynomials interpolated by these points.
Being integrated, the points are chosen as the nodes of Gauss-Lobatto quadrature for a high-order accurate approximation.
This means the approximate solution is on an element-wise orthogonal basis which furthers an intepretable analytical
approach.
The projection leads to a first-order ODE in time, the semi-discrete equation, whose
right-hand side is made up of tensor products of the discretized gradient matrices with the PDE's flux.
When the finite elements are arranged rectilinearly then these tensor products decouple into simpler products.

By casting an algorithm as tensor operations, an implementation may be accelerated by between 10-100 times
with parallelization techniques~\cite{acceleration}.
Parallelization is increasingly indispensable for scientific computing as its technology evolves towards increased
memory and processing power.
Supercomputing clusters use message-passing interface (MPI) or multiple graphics processing units (GPUs),
and most of today's desktop computers have a GPU.
On clusters, MPI-based simulations using parallelized tensor products have been run with a million processors~\cite{fisher_scale}
and research is examining MPI scaling versus parallelism on multiple GPUs~\cite{scalability}.
Yet even consumer-grade GPUs are attaining sufficient on-board memory for solution of relatively large PDEs,
with up to 24 GB of dedicated memory on the recently-released NVIDIA RTX 3000's.
This is an exciting development as it allows anyone to solve PDEs in parallel on a simple desktop machine.

As this article concerns rectilinear grids, it is also worthwhile to thoroughly review the most commonly used basis functions.
The DG method is often applied on unstructured meshes with simplex elements and the integrals needed to
obtain the gradient discretization matrices are challenging, so they are generally computed numerically~\cite{hesthaven}\cite{karniadakis}.
It was recently discovered in the context of astrophysics calculations
on a curvilinear grid that their integrals may be found in explicit
forms for the Legendre-Gauss-Lobatto (LGL) quadrature rule~\cite{teukolsky}\cite{teukolsky2}.
The LGL interpolation polynomials are found along the way to be naturally expanded in Legendre series.

This Legendre spectral expansion connects the DG method to the theory of spherical harmonics,
and the connection is useful in this work for Fourier transformation.
To understand this, first note that
Gauss-Legendre quadrature is of unit weighting because the measure of a spherical surface
projects uniformly onto an interval (its axis) due to a theorem of Archimedes~\cite{kuperberg}.
As Legendre polynomials are those harmonics invariant under rotations of the polar axis,
the quadrature shares their completeness properties, \textit{i.e.} the relation~\cite{hassani}
\begin{equation}\label{eq:spherical_completeness}
    \sum_{s=0}^{\infty}\Big(s + \frac{1}{2}\Big)P_s(x)P_s(y) = \delta(y-x),\quad\quad -1\leq x,y\leq 1.
\end{equation}
Specifically, the interpolation polynomials $\ell_j(\xi)$ through $n$ Gauss-Legendre nodes satisfy
\begin{equation}
    \ell_j(\xi) = w_j\sum_{s=0}^{n-1}\Big(s + \frac{1}{2}\Big)P_s(\xi_j)P_s(\xi),\quad\text{ and }\quad \ell_j(\xi_i) = \delta_{ij}
\end{equation}
with $\{w_j\}_{j=0}^{n-1}$, $\{\xi_j\}_{j=0}^{n-1}$ the quadrature weights and nodes respectively\footnote{Considering the
summation $\lim_{j\to\infty}\sum_{j=0}^{n-1} \ell_j(\xi)f(\xi)$ suggests that the
the interpolation polynomial plays a role in discrete integration as if it were a $\delta$-sequence on $[-1,1]$ in the
sense of distributions~\cite{stakgold}.}.
The LGL interpolant is identical to this form with a reweighted final eigenvalue $(s + 1/2)\to (s/2)$~\cite{teukolsky}.
With this relation the DG gradient matrices in the LGL basis may be found explicitly~\cite{teukolsky}\cite{teukolsky2}.

This article is structured as follows.
Section~\ref{sec:dg-proj} reviews the DG projection for PDEs and demonstrates that for a rectilinear
grid the elements may be represented as tensor products of one-dimensional elements.
Consequently only the one-dimensional discretization matrices are needed.
Using these results, Section~\ref{sec:coding} demonstrates a simple Eulerian implementation of
computing the semi-discrete equation in $n$-dimensions with GPU acceleration using upwind numerical fluxes for example.
Following this, Section~\ref{sec:example} introduces an example problem by posing a problem in plasma physics,
a magnetized plasma instability present in the Vlasov-Poisson system. Similar problems in plasma theory are
often approached by solving the Vlasov equation with DG method~\cite{rossmanith}~\cite{genia}~\cite{juno}~\cite{hakim2020}.
It should be noted that the approach advocated here will work for systems describing flows in any (reasonable) number of
dimensions. For instance by changing variables, flux, and boundary conditions in the example one can solve the
Euler equations in three dimensions.

Having posed the example problem, Section~\ref{sec:lobatto} then reviews the DG basis matrices in the Lobatto basis.
The basis analysis is then extended in Section~\ref{subsec:solving_poisson} to a Fourier spectral method used
to solve the Poisson part of Vlasov-Poisson.
Simulation results are discussed in Section~\ref{sec:simulation} with an emphasis on the computational performance
achieved using an RTX 3090 desktop GPU.
Finally, Appendix~\ref{sec:vp_ic} contains a short discussion of the initial condition used for the example problem.
All code used to produce the work shown in this article can be found on the author's GitHub at
\href{https://github.com/crewsdw/Vlasov1D2V}{https://github.com/crewsdw/Vlasov1D2V}.

\section{The DG method for rectilinear (tensor-product) grids}\label{sec:dg-proj}
Consider the conservation law for a scalar $u=u(t,x_i)$ and its flux $F^j = F^j(x_i,u)$,
\begin{equation}
  \partial_t u + \partial_j F^j = 0\label{eq:cons}.
\end{equation}
Throughout this section the up/down summation notation is used.
Equation~\ref{eq:cons} also describes vector-valued $u$ and higher-order derivatives reduced to first-order systems.
The DG method splits the domain $\Omega$ into finite elements $\Omega_\alpha$ with boundary $\partial\Omega_\alpha$,
and projects the weak or variational form of Eqn.~\ref{eq:cons} onto a polynomial basis of each element, resulting
in an approximate solution $\widetilde{u}^\alpha$.
A set of nodes is chosen $\{\xi_j\}_{j=0}^{n-1}$ within each element which are interpolated by the Lagrange polynomials
\begin{equation}\label{eq:lagrange}
  \ell_j(\xi) = \prod_{\substack{k=0\\ k\neq j}}^{n-1}\frac{\xi - \xi_k}{\xi_j-\xi_k}.
\end{equation}
The element basis is defined as these polynomials $\ell_j(\xi)$.
Rectilinear $d$-dimensional elements $\Omega$ are built up by tensor products of the one-dimensional line element
$\mathcal{L}$, so that $\Omega = \mathcal{L}_d\otimes\mathcal{L}_{d-1}\otimes\cdots\otimes\mathcal{L}_1$.
The node set of $\Omega$ is correspondingly a tensor product of each line's node set, so the $d$-dimensional
Lagrange functions can be factorized into products of the Lagrange polynomials in each dimension.
To avoid a proliferation of indices, define an index $\beta_i$ to range through the nodes
$\{\xi_j\}_{j=0}^{n-1}$ in the line element $\mathcal{L}_i$, and then let $\beta = \{\beta_0, \beta_1, \cdots, \beta_{d-1}\}$
be the multi-index collecting the nodal index of each dimension.
Additional discussion on ordering of a multi-index for tensor-product geometry constructions can be found in~\cite{fast_orthogonal}.
The Lagrange polynomial of $\Omega$ is then
\begin{equation}\label{eq:d_dim_lagrange}
  L_\beta(x_0, x_1, \cdots, x_d) = \ell_{\beta_0}(x_0)\ell_{\beta_1}(x_1)\cdots
\ell_{\beta_d}(x_d)
\end{equation}
with $x_i$ the i'th coordinate. In the projection of the approximation onto the basis functions $L_\beta$ the
expansion coefficients $u^{\alpha, \beta}$ may be labelled by the same multi-index,
\begin{equation}\label{eq:basis_exp}
    \widetilde{u}^{\alpha}(t,x) = \sum_{\beta} u^{\alpha,\beta}(t)L_{\beta}(x).
\end{equation}
Galerkin projection of Eq.~\ref{eq:cons} leads to inner products $\langle \cdot|\cdot\rangle_{\Omega_\alpha}$ over the basis functions.
Specifically, Eq.~\ref{eq:cons} is integrated on the domain against the basis functions $L_\gamma$,
\begin{equation}\label{eq:part_1}
    \langle\partial_{t}u|L_\gamma\rangle_{\Omega_\alpha} + \langle\partial_{j}F^j|L_\gamma\rangle_{\Omega_\alpha} = 0,
\end{equation}
and the basis expansion Eq.~\ref{eq:basis_exp} is substituted.
The resulting projected integral form is
\begin{equation}\label{eq:part_2}
    \frac{du^{\alpha,\beta}}{dt}\langle L_{\beta}|L_{\gamma}\rangle_{\Omega_\alpha} +
\langle \partial_jF^{\alpha,j}|L_{\gamma}\rangle_{\Omega_\alpha} = 0.
\end{equation}
Information must be passed between elements $\Omega_\alpha$ in the form of a flux, or else the elements would decouple
and the discretization would not be consistent.
This is done by integrating the flux term $\langle \partial_jF^{\alpha,j}|L_{\gamma}\rangle_{\Omega_\alpha}$ by parts
and taking the boundary term to be a function of both the local state and that of the neighbor, \textit{i.e.}
$F^\alpha|_{\partial\Omega_\alpha} \equiv \mathcal{F}(\partial\Omega_\alpha^-, \partial\Omega_\alpha^+)$ where $\partial\Omega_\alpha^-$
is the state interior to the element and $\partial\Omega_\alpha^+$ the state exterior.
The function $\mathcal{F}$, termed numerical flux, is chosen so that Eq.~\ref{eq:cons} is discretized consistently.
The requirement for consistency means $\mathcal{F}$ is related to the nature of information propagation in the system.
The single integration by parts results in the DG weak form, and one further integration by parts in the strong form,
given respectively by
\begin{align}
    \frac{du^{\alpha,\beta}}{dt}\langle L_{\beta}|L_{\gamma}\rangle_{\Omega_\alpha} &=
    F^{\alpha,\beta,j}\langle L_\beta|\partial_j L_{\gamma}\rangle_{\Omega_\alpha} -
    \mathcal{F}^{\alpha,\beta,j}\langle L_\beta|L_{\gamma}\rangle_{\partial\Omega^j_\alpha},\\
    \frac{du^{\alpha,\beta}}{dt}\langle L_{\beta}|L_{\gamma}\rangle_{\Omega_\alpha} +
    F^{\alpha,\beta,j}\langle\partial_j L_\beta|L_{\gamma}\rangle_{\Omega_\alpha} &=
    (F^{\alpha,\beta,j} - \mathcal{F}^{\alpha,\beta,j})\langle L_\beta|L_{\gamma}\rangle_{\partial\Omega^j_\alpha}
\end{align}
These inner product matrices are termed the
mass $M$, face mass $\Gamma$, advection $\mathcal{A}$, and stiffness $S$ matrices by analogy with continuum mechanics,
and their short-hand definitions are
\begin{align}
  M^\alpha_{\beta\gamma} &\equiv \langle L_\beta|L_\gamma\rangle_{\Omega_\alpha},\quad\quad
  \Gamma^{\alpha, j}_{\beta\gamma} \equiv \langle L_\beta| L_{\gamma}\rangle_{\partial\Omega^j_\alpha},\label{eq:matrices1}\\
  \mathcal{A}_{\beta\gamma}^{\alpha, j} &\equiv \langle L_\beta|\partial_j L_\gamma\rangle_{\Omega_\alpha},\quad
  S_{\beta\gamma}^{\alpha, j} \equiv \langle\partial_j L_\beta| L_\gamma\rangle_{\Omega_\alpha}.\label{eq:matrices2}
\end{align}
Now solving for the expansion coefficients results in element-wise operators composed of the products
\begin{align}
  \Upsilon^{\alpha,\beta}_{\gamma,j} \equiv (M^{-1})^{\alpha,\beta\epsilon}\mathcal{A}_{\gamma\epsilon,j}^\alpha,\quad\quad
  \Xi^{\alpha,\beta}_{\gamma,j} \equiv (M^{-1})^{\alpha, \beta\epsilon}\Gamma^{\alpha}_{\gamma\epsilon,j},\quad\quad
  D^{\alpha,\beta}_{\gamma,j} \equiv (M^{-1})^{\alpha, \beta\epsilon}S_{\gamma\epsilon,j},
\end{align}
resulting in the weak and strong form semi-discrete equations per element,
\begin{align}
  \frac{d}{dt}u^{\alpha,\beta} &= \Upsilon^{\alpha,\beta}_{\gamma,j}F^{\alpha,\gamma,j} -
  \Xi^{\alpha,\beta}_{\gamma,j}\mathcal{F}^{\alpha,\gamma,j}\label{eq:weak}\\
  \frac{d}{dt}u^{\alpha,\beta} + D^{\alpha,\beta}_{\gamma,j}F^{\alpha,\gamma,j} &=
  \Xi^{\alpha,\beta}_{\gamma,j}(F^{\alpha,\gamma,j} - \mathcal{F}^{\alpha,\gamma,j}).\label{eq:strong}
\end{align}
Considering the weak form, Eqn.~\ref{eq:weak}, the first term represents fluxes due to internal degrees of freedom within
an element, and the second term boundary fluxes.
The strong form operator $D^{\alpha,\beta}_{\gamma,\mu}$ is a direct gradient discretization called the derivative matrix,
while the operator $\Upsilon^{j,\alpha}_{k,\mu}$ approximates the gradient in integral form.
In both cases $\Xi^{\alpha,\beta}_{\gamma,\mu}$ discretizes the surface integrals between elements.
Explicitly, the multi-index $\alpha$ denotes the elements $\Omega_\alpha$,
$\beta$ the element's nodes, and $j$ the coordinates.

The basis of element $\Omega_\alpha$ is related to a reference element by an isoparametric transform and its Jacobian $J^{\alpha}_{ij}$,
so that only one set of matrices $\{\Upsilon^\beta_{\gamma,j}, \Xi^{\beta}_{\gamma,j}, D^\beta_{\gamma,j}\}$
need be determined.
These reference operators are then related to element-wise ones by $\Upsilon^{\alpha,\beta}_{\gamma,\nu} =
J^{\mu,\alpha}_{\nu}\Upsilon^{\beta}_{\gamma,\mu}$.
The following will consider operations with only the reference operators by taking elements to have an identical Jacobian.
Equations~\ref{eq:weak} \&~\ref{eq:strong} represent ODEs whose right-hand side is given by a tensor product contracting
the multi-index $\gamma$ and the coordinates $j$.
The following section shows that for rectilinear elements these contractions simplify into a sum over products with
the one-dimensional matrices $\Upsilon, D, \Xi$.

\subsection{Constituent reduction for rectilinear (tensor-product) elements}\label{subsec:reduction}
\begin{figure}[ht]
    \includegraphics[width=\textwidth]{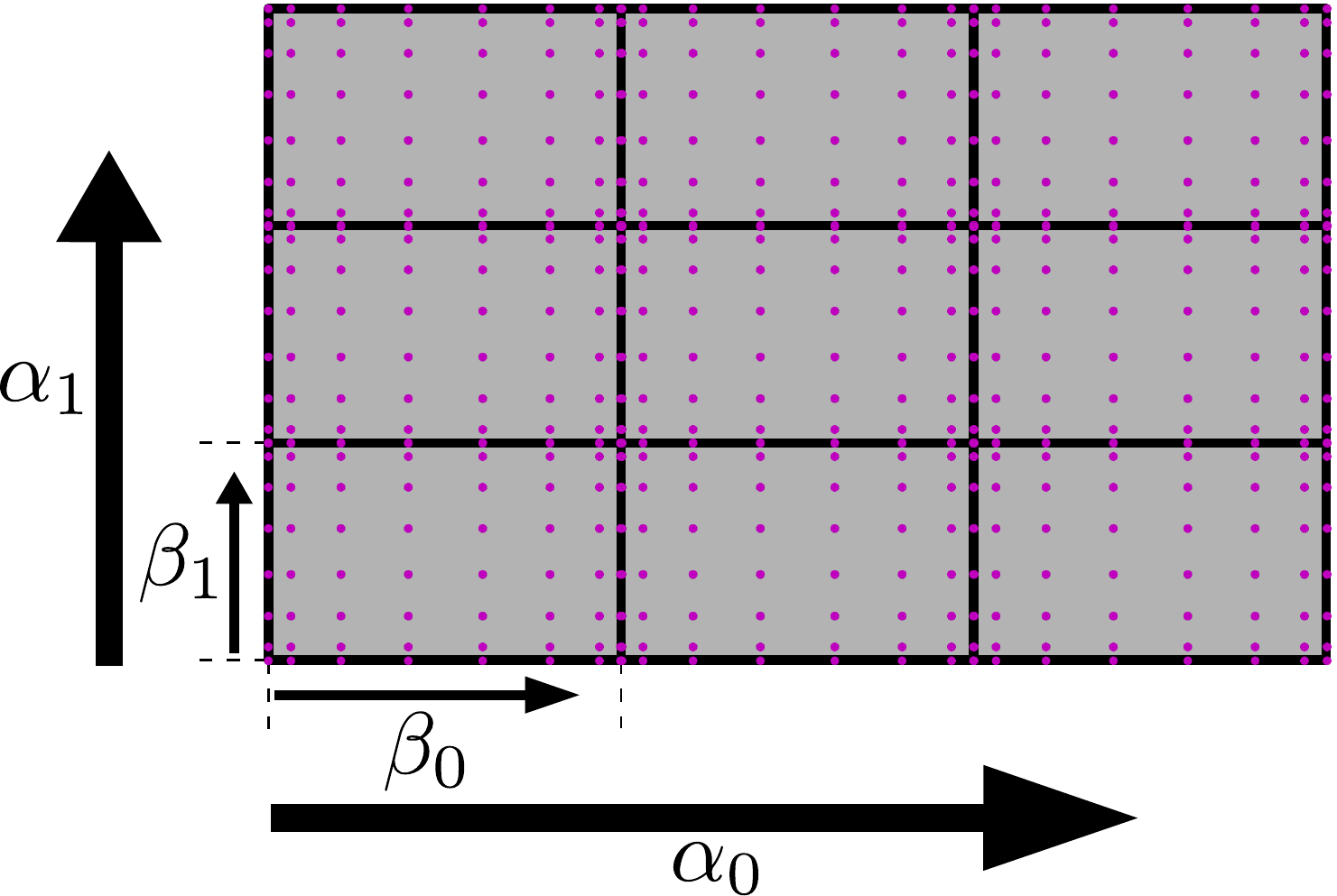}
    \caption{Rectilinear elements have a natural tensor product structure which is illustrated by the use of multi-indexes.
    Shown is a two-dimensional example with an eight-node Legendre-Gauss-Lobatto (LGL) node set, where the elements in
    directions $(0, 1)$ are indexed by $\alpha_0, \alpha_1$ and the nodal sub-element positions by $\beta_0, \beta_1$.
    Any given node's index consists of its tuple $(\alpha_0, \beta_0, \alpha_1, \beta_1)$.}\label{fig:multi_index}
\end{figure}

Now consider rectilinear $n$-cube elements specifically, with the multi-index $(\alpha,\beta)$ as shown in Fig.~\ref{fig:multi_index}.
Substitution of the factorized basis polynomials defined by Eq.~\ref{eq:d_dim_lagrange} into Eqs.~\ref{eq:matrices1}-\ref{eq:matrices2}
shows the resulting object to be composed of products of the lower-dimensional constituents.
Specifically, the mass matrix $M_{\beta,\gamma}$ with sub-element nodal multi-indices $\beta,\gamma$ decomposes into tensor
products as
\begin{equation}
  M_{\beta\gamma} = M_{\beta_{d-1},\gamma_{d-1}}\otimes M_{\beta_{d-2},\gamma_{d-2}}\otimes\cdots\otimes M_{\beta_0,\gamma_0}.\label{eq:tensor_mass}
\end{equation}
so that it may be termed the mass tensor for $d>1$.
It follows that the inverse mass tensor is $M^{-1}_{\beta\gamma} = M^{-1}_{\beta_{d-1},\gamma_{d-1}}\otimes M^{-1}_{\beta_{d-2},
\gamma_{d-2}}\otimes\cdots\otimes M^{-1}_{\beta_0,\gamma_0}$.
Similarly, the advection tensor $\mathcal{A}_{\beta\gamma}^j = \langle \partial_j L_\beta |
L_\gamma\rangle_{\Omega_\alpha}$ has $d$ components (ranging over index $j$) each given by the tensor products
\begin{equation}
  \mathcal{A}_{\beta\gamma}^j = \begin{bmatrix}M_{\beta_{d-1},\gamma_{d-1}}\otimes M_{\beta_{d-2},\gamma_{d-2}}\otimes\cdots\otimes
  \mathcal{A}_{\beta_0,\gamma_0}\\
  \vdots\quad\quad\quad\quad \vdots\quad\quad\quad \vdots\\
  M_{\beta_{d-1},\gamma_{d-1}}\otimes \mathcal{A}_{\beta_{d-2},\gamma_{d-2}}\otimes\cdots\otimes M_{\beta_0,\gamma_0}\\
  \mathcal{A}_{\beta_{d-1},\gamma_{d-1}}\otimes M_{\beta_{d-2},\gamma_{d-2}}\otimes\cdots\otimes M_{\beta_0,\gamma_0}\end{bmatrix}\label{eq:tensor_advection}
\end{equation}
with analogous results for the face mass $\Gamma$ and stiffness matrices $S$.
The components corresponding to the sub-indices of the multi-indices $\beta,\gamma$ are seen to be
the lower-dimensional matrices themselves.
Due to the tensor product mixed-product property $(A\otimes B)(C\otimes D) = (AC)\otimes (BD)$, the internal flux tensor
$\Upsilon_{\mathcal{D}}^\mu = M^{-1}_{\mathcal{D}}A^\mu_{\mathcal{D}}$ has the $d$ components,
\begin{equation}
  \Upsilon^j_{\beta\gamma} = \begin{bmatrix}I_{\beta_{d-1},\gamma_{d-1}}\otimes I_{\beta_{d-2},\gamma_{d-2}}\otimes\cdots\otimes
  \Upsilon_{\beta_0,\gamma_0}\\
  \vdots\quad\quad\quad\quad \vdots\quad\quad\quad \vdots\\
  I_{\beta_{d-1},\gamma_{d-1}}\otimes \Upsilon_{\beta_{d-2},\gamma_{d-2}}\otimes\cdots\otimes
  I_{\beta_0,\gamma_0}\\
  \Upsilon_{\beta_{d-1},\gamma_{d-1}}\otimes I_{\beta_{d-2},\gamma_{d-2}}\otimes\cdots\otimes
  I_{\beta_0,\gamma_0}\end{bmatrix},\label{eq:internal_flux}
\end{equation}
each consisting of $(d-1)$ tensor products with the identity.
The numerical flux tensor $\Xi^j_{\beta\gamma}$ follows an identical pattern, with
component $k$ as $I_{\beta_{d-1},\gamma_{d-1}}\otimes I_{\beta_{d-2},\gamma_{d-2}}
\otimes \cdots \otimes \Xi_{\beta_k,\gamma_k} \otimes \cdots\otimes I_{\beta_0,\gamma_0}$.

Equation~\ref{eq:internal_flux} means that the flux of direction $j$ only needs to be contracted with the corresponding
one-dimensional matrix of direction $j$.
To illustrate, consider the case of a $d=3$ flux $F^j$ with a rectilinear discretization.
Separate the element multi-index $\alpha$ into directional indices $a, b, c$ and $\beta$ into sub-element nodal indices $p, q, r$
so that $F^\alpha_{j,\gamma} \equiv F^{abc}_{j,pqr}$.
The internal flux product is then
\begin{align}
	\Upsilon^{\gamma,j}_{\beta}F^{\alpha}_{j,\gamma} &\equiv (I_2)^p_n(I_1)^q_m(\Upsilon_0)^r_\ell F^{abc}_{0,pqr} +
    (I_2)^p_n(\Upsilon_1)^q_m(I_0)^r_\ell F^{abc}_{1,pqr} + (\Upsilon_2)^p_n(I_1)^q_m(I_0)^r_\ell F^{abc}_{0,pqr}\\
	&= (\Upsilon_0)^r_\ell F^{abc}_{0,nmr} + (\Upsilon_1)^q_m F^{abc}_{1,nq\ell} +
    (\Upsilon_2)^p_n F^{abc}_{2,pm\ell}\label{eq:decoupled}
\end{align}
by carrying through each product with the identity.
By similar reasoning, having formed a numerical flux array $\mathcal{F}$ for the two faces of each
sub-element axis, the boundary flux product is simply
\begin{equation}
\Xi^{j,\gamma}_{\beta}\mathcal{F}^\alpha_{j,\gamma} = (\Xi_0)^{\gamma_r}_\ell\mathcal{F}^{abc}_{0,nm\gamma_r} +
(\Xi_1)^{\gamma_q}_m\mathcal{F}^{abc}_{1,n\gamma_q r} +
(\Xi_2)^{\gamma_p}_n\mathcal{F}^{abc}_{2,\gamma_p m\ell}\label{eq:numerical_flux}
\end{equation}
where $\gamma_i = \delta_{0i} + \delta_{i(n-1)}$ picks out the boundary nodes of the sub-element index.

Equation~\ref{eq:decoupled} demonstrates that orthogonal elements reduce the number of calculations needed to evaluate
the right-hand side of the semi-discrete equation.
The gain increases with dimension versus using full basis tensors.
With full tensors, $\Upsilon$ is a $\mathbb{R}^{d\times N\times N}$ array
for $N\approx n^d$ the total number of nodes in the element and $n$ the nodes per sub-element.
In contrast, with a rectilinear discretization only the $d$ one-dimensional matrices $\Upsilon\in\mathbb{R}^{n\times n}$
are needed.
Unless a non-orthogonal mesh is required for particular domain geometry, an orthogonal grid provides significant
simplification.

In some cases non-orthogonal elements are desired in certain dimensions of a problem while other dimensions are free
to use orthogonal discretization~\cite{ho_datta}.
For example the collisionless Boltzmann equation is an advection equation in $d=(D + V)$-dimensional phase space
consisting of $D$ configurational (spatial) dimensions and $V$ velocity dimensions.
As irregular boundaries will occur only in spatial dimensions, one is free to discretize velocity space orthogonally.
In this case a blended element basis $E$ may be built up from tensor products of the simplex element basis $\Delta$
and $V$ copies of the line element $\mathcal{L}_i$, \textit{i.e.} for $D = V = 2$ one has $E \equiv \Delta \otimes\mathcal{L}_1\otimes\mathcal{L}_2$.
In this case
\begin{equation}
\Upsilon_E^j = \begin{bmatrix}I_2\otimes I_1\otimes \Upsilon_\Delta\\
I_2\otimes \Upsilon_1\otimes I_\Delta\\
\Upsilon_2\otimes I_1\otimes I_\Delta\end{bmatrix}\label{eq:prism}
\end{equation}
and similarly for the numerical flux array.
The multi-index is naturally composed of unstructured spatial element information,
velocity-space grid indices $a,b$, and sub-element nodal indices $p,q$.
Equation~\ref{eq:decoupled} then generalizes to the blended case.

\section{GPU implementation with a high-level language}\label{sec:coding}
\begin{figure}[b]
    \centering
    \includegraphics[width=0.75\textwidth]{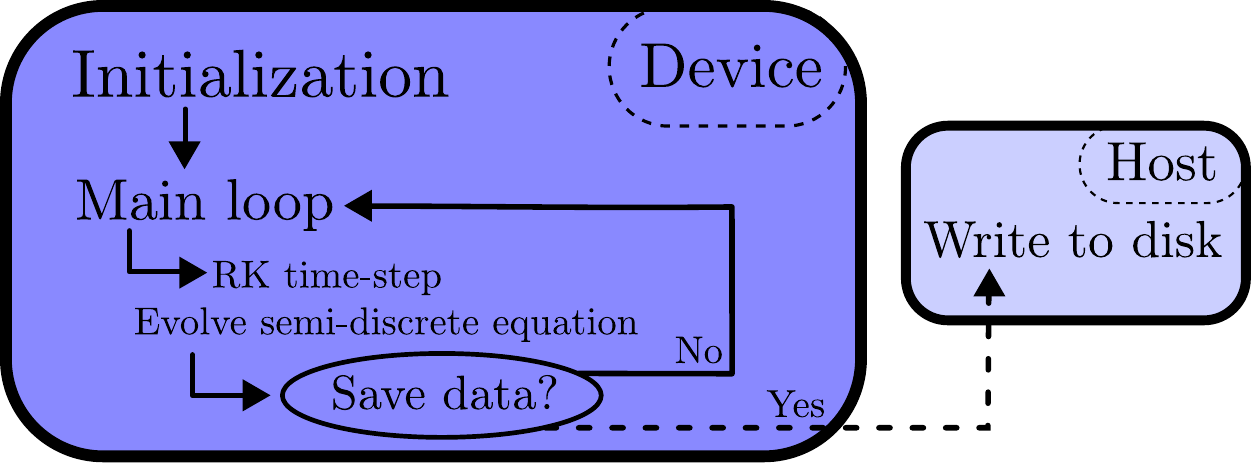}
    \caption{For performance of a parallelized DG implementation it's critical that all operations in the main time-stepping
    loop occur on the device (GPU). Device-host memory transfer should be minimized to the greatest extent possible.
    Transferring the array is only necessary at whichever time increments data is to be saved to the disk.}\label{fig:device_host}
\end{figure}

This section details an efficient implementation of the DG method 
using GPU-accelerated Python libraries for CUDA devices.
Any language which can efficiently compute tensor products of large arrays can implement DG well;
this example utilizes Python and the CuPy library~\cite{cupy_learningsys2017}.
This library mimics NumPy data array structures and may be programmed in only a few
lines of code to launch optimized kernels at execution for the tensor products appearing in the DG
semi-discrete equations.
Whichever library is used, its operations should occur entirely on device memory so that CPU-GPU transfers are only
needed to saving data.
Such a GPU-centered DG implementation is shown schematically in Fig~\ref{fig:device_host}.
Finally, the DG basis matrices are calculated via the Eqs.~\ref{eq:num_flux_wow} and~\ref{eq:internal_flux_wow}
in Section~\ref{sec:lobatto} using
i) a scientific library with Legendre polynomials such as SciPy, and
ii) tables of Legendre-Gauss-Lobatto quadrature nodes and weights available online, for example at~\cite{lobatto:calculator}.

\subsection{Evaluating the semi-discrete equation with tensor products}\label{subsec:semi_discrete_with_python}
As discussed in Section~\ref{sec:dg-proj}, when solving the semi-discrete equation on an orthogonal grid one need
only contract the flux $F^{\alpha,\beta,j}$ of each direction $j$ with its corresponding sub-element index.
With a NumPy-like data array structure this is accomplished by structuring the array data
$u^{\alpha, \beta}$ by its natural tensor-product structure.
As in Fig.~\ref{fig:multi_index}, each pair of indices for elements $\alpha_j$ and nodes $\beta_j$ corresponds to the
direction $j$, so a natural array ordering is
\begin{texttt}
    u[$\alpha_0$, $\beta_0$, $\alpha_1$, $\beta_{1}$, $\cdots$, $\alpha_{d-1}$, $\beta_{d-1}$]
\end{texttt}
and fluxes $F^{\alpha,\beta,j} = F^j(u)$ index in the same manner per dimension $j$.

Structuring the data array by both element and sub-element indices allows the semi-discrete equation to be elegantly
programmed using GPU-accelerated tensor product operations such as \texttt{cupy.tensordot(a, b, axes=)}, which compiles
a just-in-time GPU kernel to compute a tensor product with contraction on specified axes like NumPy's
\texttt{tensordot} function.
For example, consider evaluating the internal flux product $\Upsilon^{\beta}_{\gamma,j}F^{\alpha,\gamma,j}$
in the DG weak-form of Eqn.~\ref{eq:weak}.
As the product decouples in an orthogonal discretization (Eqn.~\ref{eq:decoupled}) it may be computed as a sum of tensor
products for each direction using the \texttt{basis\_product} function,
\begin{verbatim}
    def basis_product(flux, basis_matrix, axis, permutation):
        return cupy.transpose(cupy.tensordot(flux, basis_matrix,
                              axes=([axis], [0])), axes=permutation)
\end{verbatim}
where \texttt{axis} gives the sub-element index of the direction of flux,
and the function \texttt{transpose} returns a view to the original index order according to the tuple \texttt{permutation}
as \texttt{tensordot} concatenates the tensor indices remaining after contraction of the specified axes.

\subsubsection{Computing numerical fluxes}\label{subsubsec:numerical_fluxes}
The function \texttt{basis\_product} is also used to globally compute the numerical flux product
$\Xi^{\beta}_{\gamma,j}\mathcal{F}^{\alpha,\gamma,j}$, by first arranging a numerical flux array $\mathcal{F}^{\alpha,\gamma,j}$
according to the tensor-product index ordering.
The form of the function $\mathcal{F}$ depends on the chosen numerical flux scheme.
The upwind method, where $\mathcal{F}$ is a function of only one side of the interface,
is well-suited for hyperbolic problems as it solves the underlying interface, or Riemann, problem~\cite{finite_volume}.
As a unidirectional numerical flux the upwind method is also simple to adapt for the alternating fluxes used for
parabolic problems~\cite{cockburn3}.

The idea behind upwind flux is for information to flow across an element's boundaries in accordance with the PDE's
dispersion relation.
For example, in the problem $u_t + Vu_x=0$ the dispersion relation is $\omega - Vk = 0$ and information flows
to the right.
Therefore information should be sent from left to right, and numerical flux would always be chosen as depending on
values from the left.

Let $R$ stand for right, and $L$ stand for left.
Defining auxiliary variables $a^R = \text{max}(a,0)$ and $a^L=\text{min}(a,0)$
only one of which is nonzero, the upwind flux for $F=au$ with advection speed $a$ is
\begin{equation}
    \mathcal{F}(u^R, u^L) = a^{R}u^L + a^{L}u^R.
\end{equation}
An element-wise implementation of the upwind method on GPU, a bit more lengthy than the internal flux product,
is discussed in Appendix~\ref{sec:upwind_appendix}.


\section{An example problem: the Vlasov-Poisson system}\label{sec:example}
A representative problem to illustrate the DG implementation is the Vlasov-Poisson system of plasma physics.
These equations describe a probability distribution of charged particles evolving under a phase space transport equation,
self-consistently coupled through its zeroth moment to Gauss's law for the electric field as represented in potential form
by Poisson's equation.
The system is widely used in the study of low-frequency waves in hot plasmas~\cite{stix} and is often
the starting point for derivation of reduced models describing turbulent plasma dynamics such as the
gyrokinetic and quasilinear approximations~\cite{balescu}.
For a distribution of electrons $f(\bm{x}, \bm{v}, t)$ interacting through the electric potential $\Phi(x)$ in the
presence of a fixed, neutralizing background, the system
consists of
\begin{align}
    \partial_{t}f &+ \bm{v}\cdot\nabla_x f + \frac{e}{m}(\nabla_x\Phi - \bm{v}\times\bm{B}_{\text{ext}})\cdot\nabla_v f
    = 0\label{eq:vp_vlasov}\\
    \nabla^2\Phi &= \frac{en_0}{\epsilon_0}\Big(\int_{-\infty}^\infty fdv - 1\Big)\label{eq:vp_poisson}
\end{align}
where $e, m$ represent electron charge and mass respectively, $\bm{B}_{ext}$ is an external magnetic field, and $n_0$ is a reference
particle density normalizing the distribution $f$.

A useful model problem to benchmark an implementation of the system in one spatial and two velocity dimensions, \textit{i.e.}
$f(\bm{x},\bm{v}) = f(x, u, v)$, is the instability of a loss-cone distribution to perpendicular-propagating
cyclotron-harmonic waves~\cite{genia}.
By normalizing length to the Debye length $\lambda_D$, time to the plasma frequency $\omega_p^{-1}$,
velocities to the thermal velocity $v_{th} = \lambda_D\omega_p$, and the fields by setting $E_0 = v_{th}B_0$ such that
the reference plasma-cyclotron frequency ratio is unity $(\omega_p / \omega_c)_0 = 1$, the resulting normalized system is
given by the equations
\begin{align}
    \partial_{t}f &+ F^j\partial_j f = 0,\label{eq:vlasov}\\
    \partial_{xx}\Phi &= \int_{-\infty}^\infty fdv - 1,\label{eq:poisson}\\
    F &=
    \begin{bmatrix}
        u,&
        \partial_x\Phi - vB_{\text{ext}},&
        uB_{\text{ext}}
            \end{bmatrix}^T.
\end{align}
Here the normalized external magnetic field $B_{\text{ext}}$ is set such that the desired plasma-cyclotron frequency
ratio is $\omega_p / \omega_c = B_\text{ext}^{-1}$ in normalized units.

    \subsection{Initial condition for the Vlasov-Poisson system}\label{sec:vp_ic}
    The following subsection is particular to plasma physics, and the uninitiated reader should skip to the end.
    To determine an appropriate initial condition, the fundamental modes of Eqs.~\ref{eq:vp_vlasov} \&~\ref{eq:vp_poisson}
are determined by their linearization about zero-order cyclotron orbits followed by Fourier transformation in the
polar velocity coordinates $(u,v) \to (v_\perp, \varphi)$ to yield the linearized solution~\cite{stix}~\cite{gurnett},
    \begin{equation}\label{eq:linearized_solution}
        f_1(\omega,k,v_\perp,\varphi) = \Phi(\omega, k)e^{ikv_\perp\sin(\varphi)}\Big(
    \sum_{n=-\infty}^{\infty}\frac{n}{n-\omega}J_n(kv)e^{-in\varphi}\Big)
    \frac{1}{v_\perp}\frac{\partial f_0}{\partial v_\perp}
    \end{equation}
    with frequency $\omega$ normalized to the cyclotron frequency $\omega_c$ and wavenumber to the Larmor radius
$r_L = v_t/\omega_c$.
    Combination of the zeroth moment $\int_{-\infty}^\infty f_1dv$ together with the Fourier-transformed
Poisson equation yields the Harris dispersion relation, here written in the integral form~\cite{genia}~\cite{tataronis}
    \begin{equation}\label{eq:harris}
        \epsilon(\omega, k) \equiv 1 + \Big(\frac{\omega_p}{\omega_c}\Big)^2\frac{1}{\sin(\pi\omega)}\int_0^{\pi}\sin(\theta)\sin(\omega\theta)
    \mathcal{H}_0[f_0](\lambda(\theta))d\theta = 0
    \end{equation}
    where $\mathcal{H}_0[f](q)$ is the zero-order Hankel transform of $f(v)$ and the parameter $\lambda \equiv 2k \cos(\frac{1}{2}\theta)$.
    A zero-order function commonly used to model loss-cone-like probability distributions is the function
    \begin{equation}\label{eq:loss_cone}
        f_0(x, v_\perp, \varphi) = \frac{1}{2\pi\alpha^2j!}\Big(\frac{v_\perp^2}{2\alpha^2}\Big)^j\exp\Big(-\frac{v_\perp^2}{2\alpha^2}\Big)
    \end{equation}
    with ring parameter $j$ and radially-normalized gradient,
    \begin{equation}
        v_\perp^{-1}\partial_{v_\perp}f_0 = \frac{1}{v_\perp^2}\Big(j - \frac{v_\perp^2}{2\alpha^{2}}\Big)f_0.
    \end{equation}
    Equation~\ref{eq:loss_cone} describes a ring distribution, and its thermal properties are summarized in Appendix~\ref{sec:therm_appendix}.
    From the Fourier-multiplier property $\mathcal{H}_0[(v^2)^nf(v)](q) = (-\nabla^2_q)^n\mathcal{H}_0[f(v)](q)$, and the fact
    that the coefficients of successive radial Laplacians of the polar Gaussian are the Laguerre polynomials $L_j(x)$, it follows that the transform
    of Eqn.~\ref{eq:loss_cone} is given by
    \begin{equation}\label{eq:transform}
        \mathcal{H}_0[f_0](\lambda) = L_j\Big(\frac{1}{2}\alpha^2\lambda^2\Big)\exp\Big(-\frac{1}{2}\alpha^2\lambda^2\Big).
    \end{equation}
    The spatio-temporal modes of the linearized Vlasov-Poisson system are then solutions of
    \begin{equation}\label{eq:disp}
        \epsilon(\omega, k) = 1 + \Big(\frac{\omega_p}{\omega_c}\Big)^2\frac{1}{\sin(\pi\omega)}
        \int_0^{\pi}\sin(\omega\theta)\sin(\theta)L_j(\beta)e^{-\beta}d\theta = 0
    \end{equation}
    where $\beta = 2k^2\cos^2(\frac{1}{2}\theta)$.
    The dispersion function $\epsilon(\omega, k)$ may be written in closed form using hypergeometric functions
    of the form $_{2}F_{2}$.
    Yet computing them requires a power series and the trigonometric integral may be accurately computed using a quadrature method.
    A fifty-point Gauss-Legendre method was used for this work.

    Equation~\ref{eq:disp} has many solutions representing the cyclotron harmonic modes of the plasma~\cite{tataronis}.
    With ring parameter $j=6$ and frequency ratio $\omega_p / \omega_c = 10$, an unstable solution of maximum growth-rate
    occurs at $(k_0,\omega_0) \sim (0.886, 0.349i)$, for which the corresponding phase space mode is found by inverse-transforming
    Eq.\ref{eq:linearized_solution} with $\Phi(k,\omega) = \Phi_0\delta(k-k_0)\delta(\omega-\omega_0)$.
    The resulting mode can be written as a Fourier series in the cylindrical angle $\varphi$,
    \begin{align}
        f_1(x, v_\perp, \varphi) &\sim \frac{1}{v_\perp}\frac{\partial f_0}{\partial v_\perp}\text{Re}(\psi_0)\label{eq:ic}\\
        \psi_0(x, v_\perp, \varphi) &\equiv e^{i(k_0 x + k_0 v_\perp\sin(\varphi))}\sum_{n=-\infty}^{\infty}\frac{n}{n-\omega_0}J_n(k_0 v_\perp)e^{-in\varphi}
    \end{align}
    where a partial sum truncated near the approximate harmonic mode number results in a good approximation.
    For a well-converged sum this work uses an $n_{\text{max}}=20$-term approximation.

    The zeroth moment of Eq.~\ref{eq:ic} is a single harmonic $(\delta n) \sin(k_0x + \theta)$ of amplitude $\delta n$
    with some phase shift $\theta$ (though
    if desired a centered mode may be found by also considering the conjugate wavenumber's solution at $k_0'=-k_0$).
    The initial condition $f_1$ should be multiplied by a scaling factor so that the perturbation amplitude $A$
    is small enough for a valid linearization.
    Based on Eq.~\ref{eq:vlasov}, an estimate of the validity condition is $E_1 \ll v_{th}B_{\text{ext}}$.
    Combining with Poisson's equation, it follows that the condition for a valid linearization is
    \begin{equation}\label{eq:linear_valid}
        \frac{\delta n}{n_0} \ll \Big(\frac{\omega_c}{\omega_p}\Big)(k\lambda_D) \sim
        \Big(\frac{\omega_c}{\omega_p}\Big)^2(kr_L).
    \end{equation}
    This condition can be quite restrictive.
    For instance, the initial condition described above must satisfy $\delta n/n_0 \ll 0.0089$, or else a faster growing
    two stream-like instability will be observed.

\section{Element-wise operators in the Lobatto basis}\label{sec:lobatto}
This section reviews the theory of DG matrices in the semi-discrete equation
according to the method of~\cite{teukolsky}~\cite{teukolsky2}.
A review of this theory is useful as the solution is both explicit and interpretable to arbitrarily high order.
A novice reader may wish to skip to the results given in Eqs.~\ref{eq:num_flux_wow},~\ref{eq:internal_flux_wow},
and~\ref{eq:derivative}.

\subsection{Basic patterns in the Gauss-Legendre basis}\label{subsec:gauss}
It's useful to first review the result for full Gauss-Legendre quadrature as the steps are simple and
the patterns reoccur in the Lobatto basis.
The situation is simple in full Gaussian quadrature, but easy evaluation of the numerical fluxes is greatly simplified
by Lobatto quadrature by including the endpoints~\cite{hesthaven}.
Denoting the node locations by the $n$ quadrature points $\{\xi_j\}_{j=0}^{n-1}$, in addition to continuous orthogonality
the Legendre polynomials are discretely orthogonal~\cite{dunkl}~\cite{bulirsch},
\begin{equation}
    \sum_{s=0}^{n-1}w_s P_k(\xi_s)P_{\ell}(\xi_s) = \frac{1}{k+\frac{1}{2}}\delta_{k\ell}\label{eq:gauss_orthogonal}
\end{equation}
with $w_s$ the quadrature weights.
This follows from the order $2n-1$ quadrature property as the sum corresponds to the continuous integral $\int_{-1}^{1}P_k(x)P_\ell(x)dx$.
Expansions in the Lagrange basis of Eqn.~\ref{eq:lagrange} and in the Legendre basis are linearly related
due to the interpolation property $\ell_j(\xi_i) = \delta_{ij}$,
\begin{equation}
    f(\xi) = \sum_{j=0}^{n-1}f_j\ell_j(\xi) = \sum_{k=0}^{n-1}c_{k}P_k(\xi),\quad\implies\quad f_j = \mathcal{V}_j^{k}c_k\label{eq:vandermonde1}
\end{equation}
where $\mathcal{V}^k_j = P_k(\xi_j)$ is termed the (generalized) Vandermonde matrix.
The inverse transform follows directly from Eqn.~\ref{eq:gauss_orthogonal} as
$(\mathcal{V}^{-1})_j^k=w_j P_k(\xi_j)(k+\frac{1}{2})$.
In particular, the Lagrange function itself is
\begin{equation}
    \ell_j(\xi) = w_j\sum_{s=0}^{n-1}\Big(s + \frac{1}{2}\Big)P_s(\xi_j)P_s(\xi) \approx w_j\delta(\xi-\xi_j)\label{eq:lagrange_exp}
\end{equation}
making the interpolation polynomial a weighted partial sum of the completeness relation.
As the $P_j(\xi)$ are orthogonal, the mass matrix $M_{ij} = \int_{-1}^1\ell_i(\xi)\ell_{j}(\xi)d\xi
= \omega_i\delta_{ij}$ is diagonal, with inverse $M^{-1}_{ij} = \omega_i^{-1}\delta_{ij}$.
The advection and stiffness $S = \mathcal{A}^T$ matrices follow from evaluating the integrals by quadrature, \textit{e.g.}
\begin{equation}\label{eq:advection}
    \mathcal{A}_{ij} = \omega_i\ell_j'(\xi_i) = \omega_i\omega_j\sum_{s=0}^{n-1}\Big(s + \frac{1}{2}\Big)P_s(\xi_i)P_s'(\xi_j).
\end{equation}
Then the weak and strong form internal flux operators are partial sums of the delta derivative,
\begin{align}\label{eq:gradient_discretization}
    \Upsilon^i_j &= w_j\sum_{s=0}^{n-1}\Big(s + \frac{1}{2}\Big)P_s(\xi_i)P_s'(\xi_j) \approx w_j\delta'(\xi-\xi_i)\big|_{\xi=\xi_j},\\
    D^i_j &= w_j\sum_{s=0}^{n-1}\Big(s + \frac{1}{2}\Big)P_s'(\xi_i)P_s(\xi_j) \approx w_j\delta'(\xi-\xi_j)\big|_{\xi=\xi_i}, 
\end{align}
in the distributional sense.
For example, the product $\Upsilon^i_j F^j$ integrates the weak form's flux term
\begin{equation}
    \Upsilon^i_{j}F^j = \sum_{j=0}^{n-1}w_{j}F_j\sum_{s=0}^{n-1}\Big(s + \frac{1}{2}\Big)P_s(\xi_i)P_s'(\xi_j) \approx
    \sum_{j=0}^{n-1} w_j \nabla (F(\xi - \xi_i))\Big|_{\xi=\xi_j} \approx \int_{-1}^1 (\nabla F^i) d\xi
\end{equation}
by quadrature within the element.

\subsection{DG matrices in the Lobatto basis}\label{subsec:lobatto}
%
Lobatto type quadrature includes interval end-points, so that boundary information is localized to a single
interpolation basis function.
The Legendre-Gauss-Lobatto (LGL) quadrature scheme is expressed by, for $\int_{-1}^1 f(\xi)d\xi \approx \sum_{j=0}^{n-1}w_j f(\xi_j)$,
the nodes $\{\xi_j\}_{j=0}^{n-1}$ and weights $\{w_j\}_{j=0}^{n-1}$ according to
\begin{equation}
  \{\xi_j|(1 - \xi_j^2)P_{n-1}'(\xi_j)=0\},\quad\quad w_j = \frac{2}{n(n-1)}\frac{1}{(P_{n-1}(\xi_j))^{2}}.\label{eq:quad_weights}
\end{equation}
The above rule integrates polynomials of degree $\leq 2n-3$~\cite{bulirsch}.

Discrete orthogonality of classical orthogonal polynomials under Gaussian quadrature follows from that of the eigenvectors
of the defining recurrence relation represented as a tri-diagonal Jacobi matrix~\cite{dunkl}.
According to the article of Gautschi~\cite{gautschi}, the quadrature nodes and weights were often constructed via the
eigenvalues of the matrix for the Legendre recurrence relation with a modified final row and column, particularly
when using Jacobi polynomials, rather than the form of Eq.~\ref{eq:quad_weights}.
This was originally proposed in a classic work by Golub~\cite{golub}.
However, a consequence of this modified Jacobi matrix eigenvalue problem is that
the Legendre polynomials remain discretely orthogonal under LGL quadrature,
\begin{equation}
  \sum_{s=0}^{n-1}w_s P_k(\xi_s)P_\ell(\xi_s) = \gamma_k\delta_{k\ell},\quad\quad \gamma_k \equiv \begin{cases}\frac{2}{2k+1} & k< n-1,\\
  \frac{2}{n-1}& k=n-1\end{cases}\label{eq:discrete_orthogonality}
\end{equation}
with a modified final eigenvalue from $(k+\frac{1}{2})^{-1}$ to $(k/2)^{-1}$.
This also follows by direction calculation as in~\cite{teukolsky},
where $k < n-1$ follows by quadrature and $\gamma_{n-1}$ by the boundary weights $w_0,w_{n-1}$.

Just as for Gauss-Legendre quadrature, the discrete orthogonality relation expresses the coefficients of the
Vandermonde matrix and its inverse.
That is, with the same expansion of Eqn.~\ref{eq:vandermonde1} the matrix elements are $\mathcal{V}_j^k \equiv P_k(\xi_j)$
with inverse components $(\mathcal{V}^{-1})_j^k = w_j\gamma_k^{-1}P_k(\xi_j)$ following directly from Eqn.~\ref{eq:discrete_orthogonality}.
The spectral transform from the Lagrange to Legendre spectral basis is identical in form to full Gauss-Legendre quadrature.
In particular the modal form of the Lagrange basis functions themselves is of the same form,
\begin{equation}
  \ell_j(\xi) = (\mathcal{V}^{-1})_j^kP_k(\xi) = w_j\sum_{s=0}^{n-1}\gamma_k^{-1}P_s(\xi_j)P_s(\xi) \approx \omega_j\delta(\xi - \xi_j)\label{eq:int_exp}
\end{equation}
and expresses a re-weighted partial summation of the completeness theorem due to the
Lobatto weighting of the last term in the series, $\gamma_{n-1}$.

As discovered in~\cite{teukolsky}, the Lagrange function spectral form of Eqn.~\ref{eq:int_exp} reveals the Lobatto basis
mass matrix to be diagonal with a rank-one update.
This makes it explicitly invertible.
To review the quoted result, let $A_m \equiv \int_{-1}^1P_m(\xi)P_m(\xi)d\xi$.
Noting that $A_m = \frac{2}{2m+1} = \gamma_m$ of Eqn.~\ref{eq:discrete_orthogonality} for $m < n-1$, the mass matrix integrates to
\begin{align}
  M_{ij} = \int_{-1}^1\ell_i(\xi)\ell_j(\xi)d\xi &=
  \sum_{k,m=0}^{n-1}\frac{w_iw_j}{\gamma_k\gamma_m}P_k(\xi_i)P_m(\xi_j)\int_{-1}^1P_k(\xi)P_m(\xi)d\xi\\
                                                 &=
                                                 w_i\sum_{k=0}^{n-1}\frac{w_j}{\gamma_k}P_k(\xi_i)P_k(\xi_j) + \frac{A_{n-1} -
  \gamma_{n-1}}{\gamma_{n-1}^2}w_iw_jP_{n-1}(\xi_i)P_{n-1}(\xi_j)\\
                                                 &=
                                                 w_i(\delta_{ij} - \alpha w_jP_{n-1}(\xi_i)P_{n-1}(\xi_j))
\end{align}
as the first sum equals $\ell_i(\xi_j)$ and where $\alpha = (\gamma_{n-1}-A_{n-1})\gamma_{n-1}^{-2} = \frac{n(n-1)}{2(2n-1)}$.
The mass matrix is full, but differs from the diagonal Gauss-Legendre result by a rank-one matrix.
Applying the identity $(I - uv^T)^{-1} = I + \frac{uv^T}{1 - v^Tu}$ inverts the mass matrix as
\begin{equation}
  M^{-1}_{ij} = \frac{1}{w_i}\delta_{ij} + \frac{n}{2}P_{n-1}(\xi_i)P_{n-1}(\xi_j).\label{eq:minv}
\end{equation}
The DG matrices in the Lobatto basis follow from Eqn.~\ref{eq:minv}.
Fortunately, the face mass matrix $\Gamma$, advection matrix $A$ and stiffness matrix $S$ follow more easily than the
mass matrix.
For the interval $[-1,1]$ the face mass matrix $\Gamma$ is a Kronecker delta picking out the boundary nodes, while the
advection matrix $\mathcal{A}$ is maximal order $2n-3$, so by quadrature
\begin{equation}\label{eq:lobatto-advection}
  \mathcal{A}_{ij} = \langle\ell_i(\xi)|\partial_\xi\ell_j(\xi)\rangle = w_i\ell_j'(\xi_i), 
\end{equation}
with the stiffness matrix its transpose $S_{ij} = \mathcal{A}_{ij}^T = w_j\ell_i'(\xi_j)$.
Using the projection matrices $M^{-1}$, $\mathcal{A}$, and $S$, and the Lagrange function of Eqn.~\ref{eq:int_exp},
the DG operators follow as
\begin{align}
  \Upsilon^j_k &= \frac{w_k}{w_j}\ell_j'(\xi_k) + \frac{n}{2}w_kP_{n-1}(\xi_j)P_{n-1}'(\xi_k),\\
  \Xi^{j}_{\gamma^k} &= \frac{1}{w_j}\delta_{j\gamma^k} + \frac{n}{2}P_{n-1}(\xi_j)P_{n-1}(\xi_{\gamma^k}),\label{eq:num_flux_wow}\\
  D^j_k &= \ell_k'(\xi_j) 
\end{align}
Here, the numerical flux array $\Xi$ is simply the first and last columns of the inverse mass matrix where the index
$\gamma^k = \delta_{k0} + \delta_{k(n-1)}$ denotes the interval end-points.
The term $P_{n-1}'(\xi_k)$ in $\Upsilon^j_k$ expressing the rank-one difference from the Gauss-Legendre result arises
from the identity
\begin{equation}\label{eq:cool_identity}
  \sum_{s=0}^{n-1}P_{n-1}(\xi_s)\ell_s'(\xi_j) = P_{n-1}'(\xi_j)
\end{equation}
which follows from the orthogonality relation Eqn.~\ref{eq:discrete_orthogonality}.
The result for $D^j_k$ is found by accounting for boundary terms after a discrete integration by parts and canceling
the remainder.
By then applying Eqn.~\ref{eq:int_exp} the weak form operator $\Upsilon$ becomes simply
\begin{equation}
  \Upsilon_k^j = w_k\sum_{s=0}^{n-1}\Big(s+\frac{1}{2}\Big)P_s(\xi_j)P_s'(\xi_k).\label{eq:internal_flux_wow}
\end{equation}
This form of the internal flux operator clearly demonstrates it as a flux discretization, that is, as a partial
derivative of the Legendre completeness relation.
In the limit of many quadrature nodes the internal flux operator
approaches $\sim\omega_k\delta'(\xi-\xi_j)\big|_{\xi=\xi_k}$, picking out the flux gradient
as $\Upsilon^j_{k}F^k\sim \sum_k w_k\delta'(\xi-\xi_j)|_{\xi=\xi_k}F^k\approx \nabla F^j$, just as Gauss-Legendre
but of lower accuracy due to the reweighted value $\gamma_{n-1}$.
Similarly, the derivative matrix is seen to be
\begin{equation}\label{eq:derivative}
  D^j_k = w_k\sum_{s=0}^{n-1}\frac{1}{\gamma_s}P_s'(\xi_j)P_s(\xi_k)
\end{equation}
so that, through integration by parts to return to the strong form, the components differ from the completeness relation
by the last summation term~\cite{teukolsky}.
This difference alters only the boundary values of $D^j_k$ as
$P_{n-1}'(\xi_j) = 0$ except for $\xi_{0}$ and $\xi_{n-1}$. 

\subsection{Spectral methods on a discontinuous Lobatto grid: Fourier analysis}\label{sec:fourier_series}
This section discusses a complementary approach to the DG method using its discontinuous interpolation polynomial basis.
Problems can be solved by transforming to a global spectral basis, \textit{i.e.} a complete set of orthogonal functions
for the entire domain.
Since the underlying node set remains the LGL nodes this approach naturally complements the DG method.
The transformation is summarized by a tensor whose components are called connection coefficients, because they
connect different families of basis functions~\cite{fast_orthogonal}.

Global spectral methods work well for elliptic problems, as illustrated here by solving Poisson's equation.
This makes the method of interest for coupled hyperbolic-elliptic systems, with the hyperbolic part solved using DG
method and the elliptic part by the method of this section.
Other applications of this high-order Fourier transform are as a global filter for anti-aliasing, accurate
calculation of convolutions, and post-processing spectral analysis.

Note that in DG method functions are approximated by an $N$-element piecewise interpolation,
\begin{equation}\label{eq:global_approx}
    f(x) = \bigoplus_{m=0}^{N-1}\sum_{j=0}^{n-1}f_{mj}\ell_j(x)
\end{equation}
with the direct sum $\bigoplus$ over elements and $\ell_j(x)$ the Lagrange polynomial of the nodes $\{\xi_j\}_{j=0}^{n-1}$.
Now consider the determination of Eqn.~\ref{eq:global_approx}'s Fourier coefficients.
As the approximation satisfies the Dirichlet conditions, it is expandable in Fourier series as~\cite{lanczos}
\begin{equation}\label{eq:fourier_series}
    f(x) = \sum_{p=-\infty}^{\infty}c_{p}e^{ik_px},\quad\quad c_p = \frac{1}{L}\int_0^{L}f(x)e^{-ik_px}dx
\end{equation}
where $L$ is the domain length and $k_p = \frac{2\pi}{L}p$ the wavenumbers.
Because the interpolation polynomial is an approximate identity $\ell_j(\xi)\approx w_j\delta(\xi-\xi_j)$,
its Fourier transform approximates the mode of that node, $w_j e^{-ik_p\xi_j}$.
This is seen by carrying out the Fourier integral using the interpolation polynomial
expansion (Eqn.~\ref{eq:int_exp}) having applied the affine transformation
\begin{equation}
    \xi = J_m(x - \bar{x}_m)
\end{equation}
per element with $\bar{x}_m$ the mid-point of the $m$'th element, $J_m = 2 / (\Delta x)_m$ the element Jacobian,
and $(\Delta x)_m$ its width.
The function $\ell_j(\xi)$ is a combination of spherical harmonics $P_j(\xi)$, so its Fourier transform is a
combination of spherical waves
as the transformation reduces to the integral~\cite{fast_orthogonal}
\begin{equation}
    \int_{-1}^1 P_s(\xi)e^{-ik_pJ_m^{-1}\xi}d\xi = 2(-i)^{s}j_s(J_m^{-1} k_p)
\end{equation}
with $j_s(\zeta)$ the spherical Bessel function.
The solution $c_p$ is cleanly expressed by defining
\begin{align}\label{eq:spectral_transform}
    \mathbb{T}_p^{mj} \equiv&~\frac{(\Delta x)_m}{L}e^{-ik_p\bar{x}_m}w_j
\sum_{s=0}^{n-1}\gamma_s^{-1}(-i)^sP_s(\xi_j)j_s(J_m^{-1}k_p)\\
\approx&~w_j e^{-ik_px_{mj}}\frac{2(\Delta x)_m}{L}
\end{align}
such that $c_p = \mathbb{T}_p^{mj}f_{mj}$. Equation~\ref{eq:spectral_transform} is an $n$-term sum of the expansion~\cite{hassani},
\begin{equation}\label{eq:spherical_expansion}
    e^{-ikx} = 2\sum_{s=0}^{\infty}\Big(s + \frac{1}{2}\Big)(-i)^s P_s(x)j_s(k).
\end{equation}
with a Lobatto-weighted coefficient $\gamma_{n-1}$.
The transform encoded by $\mathbb{T}_p^{mj}$ depends on the grid only.

\subsubsection{Approximation capability of the transformation}
\begin{figure}[ht]
    \centering
    \includegraphics[width=0.75\textwidth]{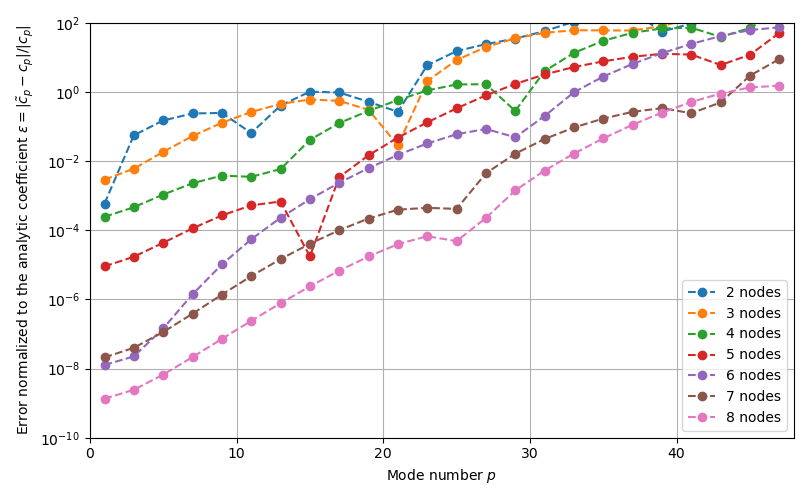}
    \caption{Normalized coefficient error $\epsilon = |\widetilde{c}_p-c_p|/|c_p|$ of the elliptic cosine on $[-1/2, 1/2]$
        using the transform $\widetilde{c}_p = \mathbb{T}_p^{mj}f_{mj}$ with $N=20$-elements.
        Even-numbered modes, not shown, float at $10^{-16}$.
    This example explores the location of the aliasing point based on the first local error minimum.
    The usual discrete Fourier transform's aliasing point (the Nyquist frequency) is at half the sampling frequency,
        here the tenth mode $p = 10$.
        The transform by $\mathbb{T}_p^{mj}$ for $n>1$ shifts the aliasing point to greater $p$.
    Based on these observations, for $n\gtrsim 6$ the Nyquist mode can be taken as roughly the number of elements,
        here $p_{\text{max}} \sim 20$.}\label{fig:errors}
\end{figure}

As the polynomial order of the basis $\ell_j(\xi)$ is increased, the spectrum is approximated by
successively higher order quadratures.
In practice the inverse-transformation of Eqn.~\ref{eq:fourier_series} should be summed up to the highest mode
$p_{\text{max}}$ 
prior to aliasing.
To diagnose the aliasing phenomenon, consider approximating the elliptic cosine $\text{cn}(x|m)$ normalized to $x\in [-1/2, 1/2]$.
Its Fourier expansion is
\begin{equation}
\text{cn}(4Kx|m) = \frac{\pi}{K\sqrt{m}}\sum_{n=0}^{\infty}\text{sech}(\pi(1+2n)K'/K)\cos(2\pi(1+2n)x)
\end{equation}
where $K(m)$ is the quarter-period (complete elliptic integral) and $K'(m)=K(1-m)$.
Examining the spectral error associated with a discontinuous interpolation as in Fig.~\ref{fig:errors}
(with $\sqrt{m}=1-10^{-6}$) shows the
number of accurate modes prior to the aliasing limit to go beyond the usual Nyquist frequency (that is, half the sampling rate)
associated with the usual discrete Fourier transform.
At high enough order ($n\gtrsim 6$), the aliasing point occurs at greater than twice the Nyquist limit
of a Fourier series on equispaced points (or piecewise constant interpolating polynomials).

Note that if all elements are of equal width, then the indices corresponding to nodes $j$ and elements $m$ factorize, and
the transform can be written much as a classic Fourier transform matrix on equidistant data.
The piecewise-constant polynomial $n=1$ corresponds with the discrete Fourier transform when the inverse-transform sum
is truncated at the Nyquist mode, as $j_0(k_p)= \text{sinc}(k_p)$ factors from the transform components and
the matrix may then be inverted by the discrete orthogonality of Fourier modes on equidistant points.
For $n>1$ the inverse transformation is not exact yet increases in accuracy with $n$, illustrated by solving Poisson's equation.
In this sense the scheme constitutes a kind of high-order discrete Fourier transform.

\subsubsection{An application: solving Poisson's equation}\label{subsec:solving_poisson}
\begin{figure}[t]
    \centering
    \begin{subfigure}{.45\textwidth}
        \centering
        \includegraphics[width=\linewidth]{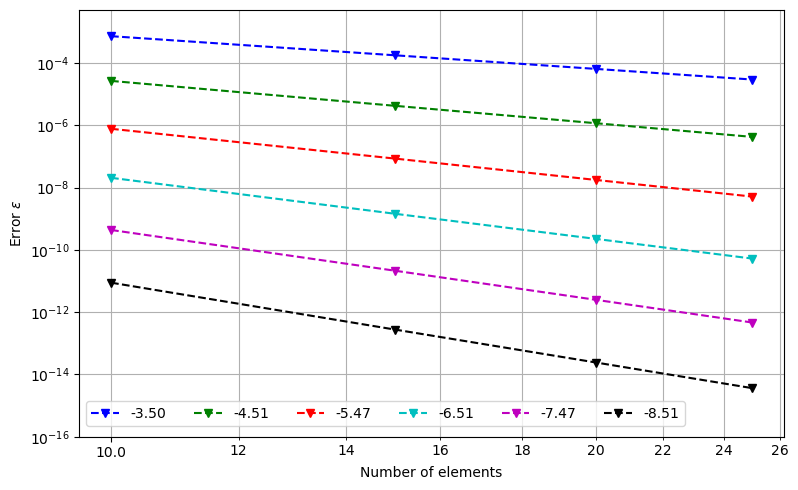}
        \caption{Error in the potential $\Phi(x)$.}
        \label{fig:sub-first}
    \end{subfigure}
    \begin{subfigure}{.45\textwidth}
        \centering
        \includegraphics[width=\linewidth]{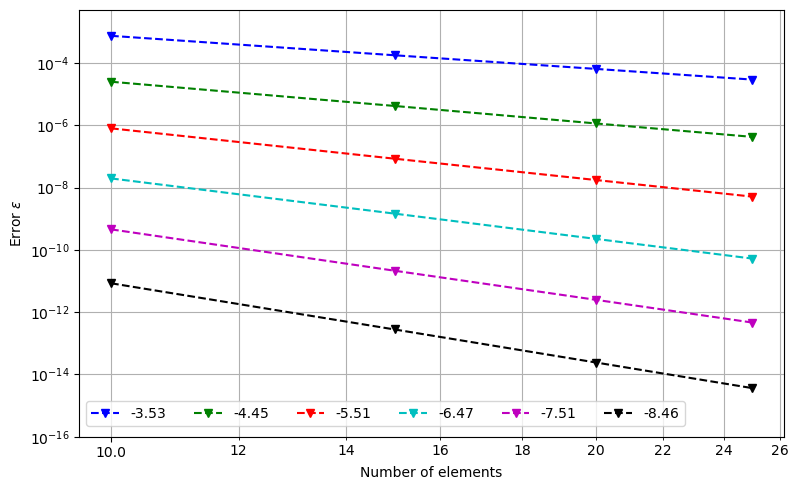}
        \caption{Error in the field $E(x)=-\frac{d\Phi}{dx}$.}
        \label{fig:sub-second}
    \end{subfigure}
    \caption{Convergence with element width $h$ and polynomial order $p$, or $hp$-refinement, for
    the spectral solution of Poisson's equation $\Phi''=\sin(2\pi x)$ using LGL nodes of $n=3$-$8$
        and $N=10$, $15$, $20$, and $25$ elements, where the legends indicate
        the line of best fit's slope sequentially for $n=3$-$8$.
    An $h$-convergence of $\mathcal{O}((\Delta x)^{n+1/2})$ is observed, the same as obtained by
    solving the DG strong form with stabilized central numerical fluxes as in~\cite{hesthaven}.}\label{fig:poisson_error}
\end{figure}

For an application of the spectral method, consider the problem
\begin{equation}\label{eq:poisson1}
    \frac{d^2\Phi}{dx^2} = f(x)  
\end{equation}
with periodic boundary conditions $\Phi(x + L) = \Phi(x)$ and $x\in [0, L]$.
A spectral solution is obtained by sampling the source $f(x)$ on the
piecewise quadrature nodes and expanding it in the form of Eqn.~\ref{eq:global_approx}.
Having obtained the Fourier coefficients $c_p = \mathbb{T}_p^{mj}f_{mj}$ by application of the spectral transform tensor (Eqn.~\ref{eq:spectral_transform}),
the solution in the spectral domain follows from the transform of Eq.~\ref{eq:poisson1},
\begin{equation}\label{eq:fourier_poisson}
    -k_p^2\widetilde{\Phi}(k_p) = c_p
\end{equation}
for a spectral solution $\widetilde{\Phi}(k) = \sum_{p=-\infty}^{\infty}-k_{p}^{-2}c_{p}\delta(k-k_p)$.
Summing modes up to $p_\text{max}$, the solution obtained is
\begin{align}\label{eq:poisson_solution}
    \Phi(x) &= \bigoplus_{m=0}^{N-1}\sum_{j=0}^{n-1}\Phi_{mj}\ell_j(x)\\
    \Phi_{mj} &= \sum_{p=-p_{\text{max}}}^{p_{\text{max}}}-k_p^{-2}c_p e^{ik_{p}x_{mj}}
\end{align}
where $x_{mj}$ is the location of the j'th node in the m'th element.
The electric field $E(x) = -\frac{d\Phi}{dx}$ is obtained in the same manner,
\begin{equation}\label{eq:electric_sol}
    E_{mj} = \sum_{p=-p_\text{max}}^{p_{\text{max}}}-((ik_p)^{-1}c_p)e^{ik_{p}x_{mj}}.
\end{equation}
Often only the field $E(x)$ is needed, like in the Vlasov-Poisson system.
The solution obtained by this high-order spectral method is observed in Fig.~\ref{fig:poisson_error} to be
$\mathcal{O}((\Delta x)^{n+1/2})$-accurate for a simple benchmark problem where the error is calculated using the broken
$\ell_2$-norm,
\begin{equation}\label{eq:broken_l2}
    \varepsilon = ||u - u_\text{exact}||_{2} = \frac{1}{N}\sum_{m=1}^{N}||u_m - u_{\text{exact}}||_{2,m},\quad
||u||_{2,m} \equiv \sqrt{\int_{x_m}^{x_m+(\Delta x)_m} u(x)^2 dx}.
\end{equation}

\begin{figure}[h]
    \centering
    \begin{subfigure}{.32\textwidth}
        \centering
        \includegraphics[width=\linewidth]{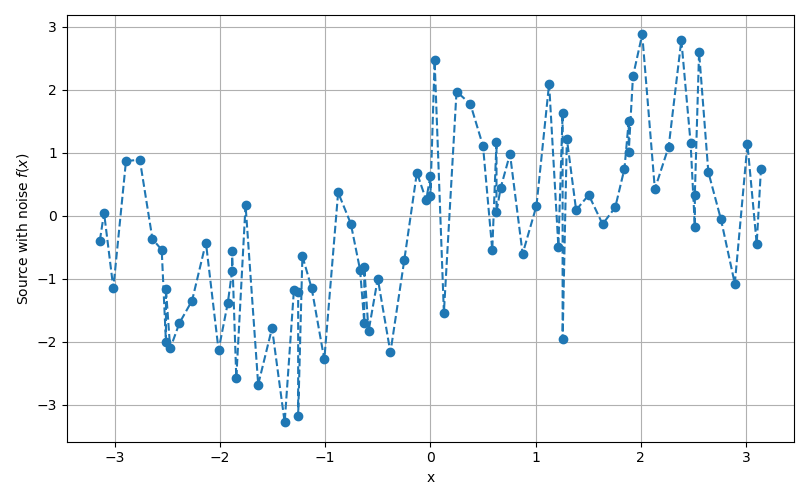}
        \caption{Noisy source function $f(x)$.}
        \label{fig:sub_poisson0}
    \end{subfigure}
    \begin{subfigure}{.32\textwidth}
        \centering
        \includegraphics[width=\linewidth]{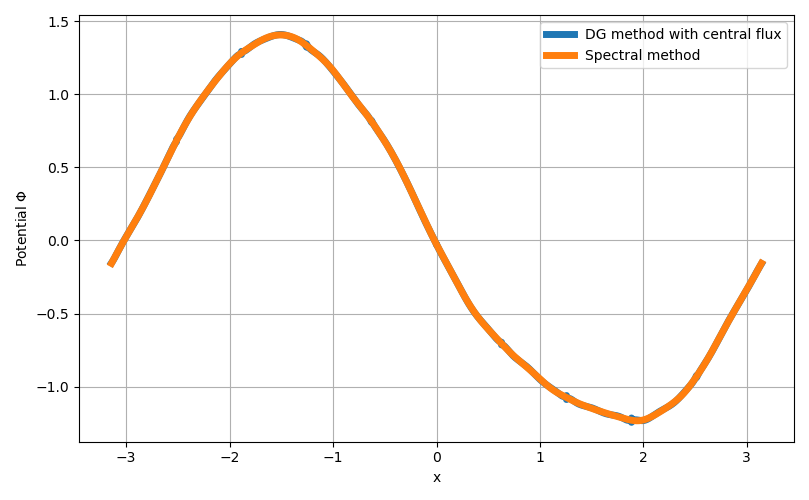}
        \caption{Potential solution $\Phi(x)$.}
        \label{fig:sub_poisson1}
    \end{subfigure}
    \begin{subfigure}{.32\textwidth}
        \centering
        \includegraphics[width=\linewidth]{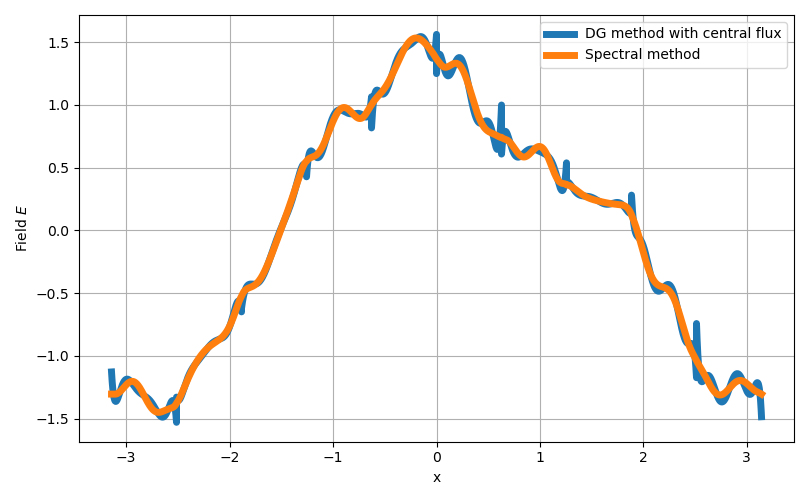}
        \caption{Field solution $E(x)=-\frac{d\Phi}{dx}$.}
        \label{fig:sub_poisson2}
    \end{subfigure}
    \caption{Comparison of solution to Poisson's equation $\nabla^2\Phi = f(x)$ with $x\in[-\pi,\pi]$ between the
    LGL-Fourier spectral method and a matrix-based method using stabilized central numerical fluxes as in~\cite{hesthaven},
        where the source density $f(x) = \sin(x) + \mathcal{N}(0,1)$
    and $\mathcal{N}(0,1)$ is a normally distributed random variable of mean $0$ and variance $1$.
    The domain consists of $N=10$ elements of $n=8$ nodes each, and Figs.~\ref{fig:sub_poisson1},~\ref{fig:sub_poisson2}
    plot the solutions using the basis functions $\bigoplus_m\sum_j y_{mj}\ell_j(x)$.}\label{fig:poisson_noise}
\end{figure}

Figure~\ref{fig:poisson_noise} studies the case of a source density $f(x)$ with discontinuities on element boundaries,
a situation admitted by the DG scheme and encountered in practice when solving coupled hyperbolic-elliptic equations, by
introducing large-amplitude random noise.
As the Fourier series is single-valued on element boundaries, the field and potential solutions via the spectral
method are $\mathcal{C}^0$.

\section{Example results: simulating the Vlasov-Poisson system}\label{sec:simulation}
This section explores numerical solutions to the Vlasov-Poisson problem posed in Section~\ref{sec:example} in order
to illustrate the DG method on GPU.
As in Section~\ref{sec:vp_ic}, the chosen initial condition consists of the velocity distribution
$f(x,u,v) = f_0(x,u,v) + \epsilon f_1(x, u, v)$ with $\epsilon$ a small parameter.
For the equilibrium distribution $f_0$, the loss cone distribution Eq.~\ref{eq:loss_cone} with ring
parameter $j=6$ and thermal velocity $\alpha = 1$ is chosen, while the perturbation mode $f_1$ is given by Eq.~\ref{eq:ic}.
The parameter $\epsilon$ is chosen small enough to satisfy the linearization condition of Eq.~\ref{eq:linear_valid},
though $\epsilon$ is not otherwise not important.
For these examples $\epsilon$ is chosen such that the perturbed density $\int_{-\infty}^\infty fdv = 0.002$.
Further, the normalized magnetic field is taken as $B_{\text{ext}}=0.1$.

To summarize the numerical method, Eq.~\ref{eq:vlasov} is a conservation law and so discretized as in Section~\ref{sec:dg-proj}.
As the Vlasov equation is a first-order hyperbolic problem the upwind numerical fluxes of
Section~\ref{subsubsec:numerical_fluxes} are used, which are also described in~\cite{hesthaven}~\cite{finite_volume}.
For time integration, the Shu-Osher third-order explicit SSP-RK method~\cite{ssprk} with spatial-order dependent CFL numbers
given in~\cite{cockburn} is used to evolve the semi-discrete equation in the weak form of Eq.~\ref{eq:weak}.
Poisson's equation, Eq.~\ref{eq:poisson}, is solved at each RK stage via the method of
Section~\ref{subsec:solving_poisson}, \textit{i.e.} the DG Fourier-spectral method.

\subsection{Two cyclotron-harmonic instability case studies}\label{subsec:two_case_studies}
\begin{figure}[h]
    \centering
    \begin{subfigure}{0.5\textwidth}
    \includegraphics[width=\textwidth]{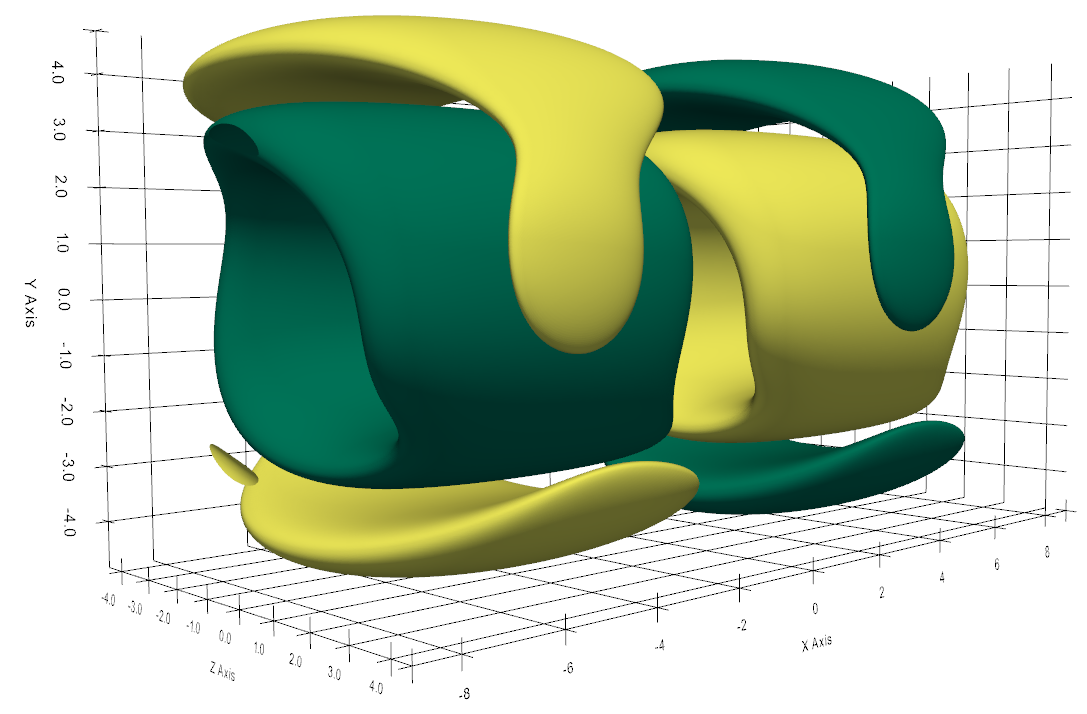}
        \caption{Mode of simulation $A$ with $\omega_r = 0$. The mode changes sign in velocity space
        where $\partial_{v_\perp}f_0 \sim 0$.}\label{subfig:mode0_ic}
    \end{subfigure}
    \hfill
    \begin{subfigure}{0.4\textwidth}
        \includegraphics[width=\textwidth]{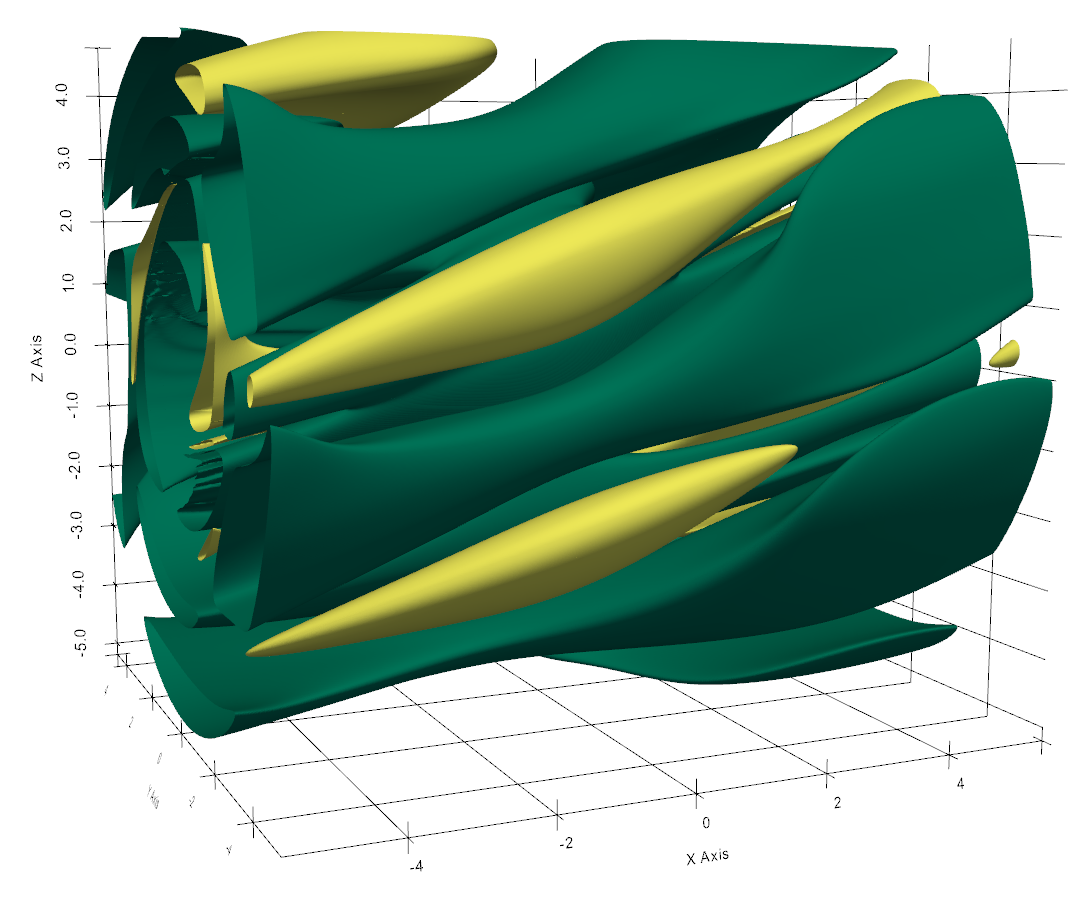}
        \caption{Simulation $B$'s perturbation corresponding to a mode with $\omega_r\neq 0$.}\label{subfig:mode1_ic}
    \end{subfigure}
    \caption{The perturbations $f_1(x,u,v)$ consist of twisting islands in the phase space,
        capturing the combined physics of translation due to particle momentum, acceleration by the electric field, and rotation
        by the magnetic field. Shown here are the perturbations $f_1 \equiv f - f_0$ at time $t=0$ for simulations $A$ and $B$,
        each with iso-surfaces at $30\%$ of the minimum (green) and maximum (yellow). The plot axes $(X,Y,Z)$ correspond
        to the phase space coordinates $(x, u, v)$.}\label{fig:perturbation}
\end{figure}

Two example simulations $A$ and $B$ are performed using as perturbation eigenvalues
the pairs $(k_A,\omega_A)\sim (0.886, 0.349i)$ and $(k_B, \omega_B)\sim (1.4, 1.182 + 0.131i)$,
each a solution of Eq.~\ref{eq:harris}, the dispersion function.
Directly exciting a mode is a general method; any solution $\{f_1, (k, \omega)\}$ of the linearized equations can be
simulated for comparison to linear theory.
The modes are chosen with $\omega_i>0$ in order to observe the transition from linear to nonlinear amplitudes
with a verifiable growth rate.
Lastly, the domain length is taken as $L = 2\pi / k$ and the domain velocity limits are set to $u_{\text{max}}, v_{\text{max}}\sim \pm 8.5$.
Figure~\ref{fig:perturbation} shows the perturbations $f_{1}(x,u,v)$ at $t=0$.

The problem is run with an element resolution $(N_x, N_u, N_v) = (25, 50, 50)$ and $n=8$ nodal basis per dimension
for a total of $32$ million nodes.
These instabilities grow on a slow time-scale relative to the plasma frequency;
that is, they grow at a fraction of the cyclotron time-scale $\omega_c^{-1}$, while time $t$ is
normalized to the plasma frequency $\omega_p^{-1}$.
Further, the run conditions have $\omega_p = 10\omega_c$.
This means that the instabilities reach their nonlinear saturation phase after many hundreds of plasma periods.
Simulation $A$ reaches saturation around $t=100$ but was run to $t=175$, while simulation $B$ saturates at around
$t=175$ and was stopped at $t=200$.
Each case requires approximately $50000$ three-stage time steps to reach the stop time,
with a machine run-time of several hours on an RTX 3090 GPU.
For a sense of magnitude, an equivalent single-threaded implementation on CPU, at least thirty times slower,
would require at least one week of calculation time.

\begin{figure}[b!]
    \centering
    \begin{subfigure}{0.32\textwidth}
        \centering
        \includegraphics[width=\textwidth]{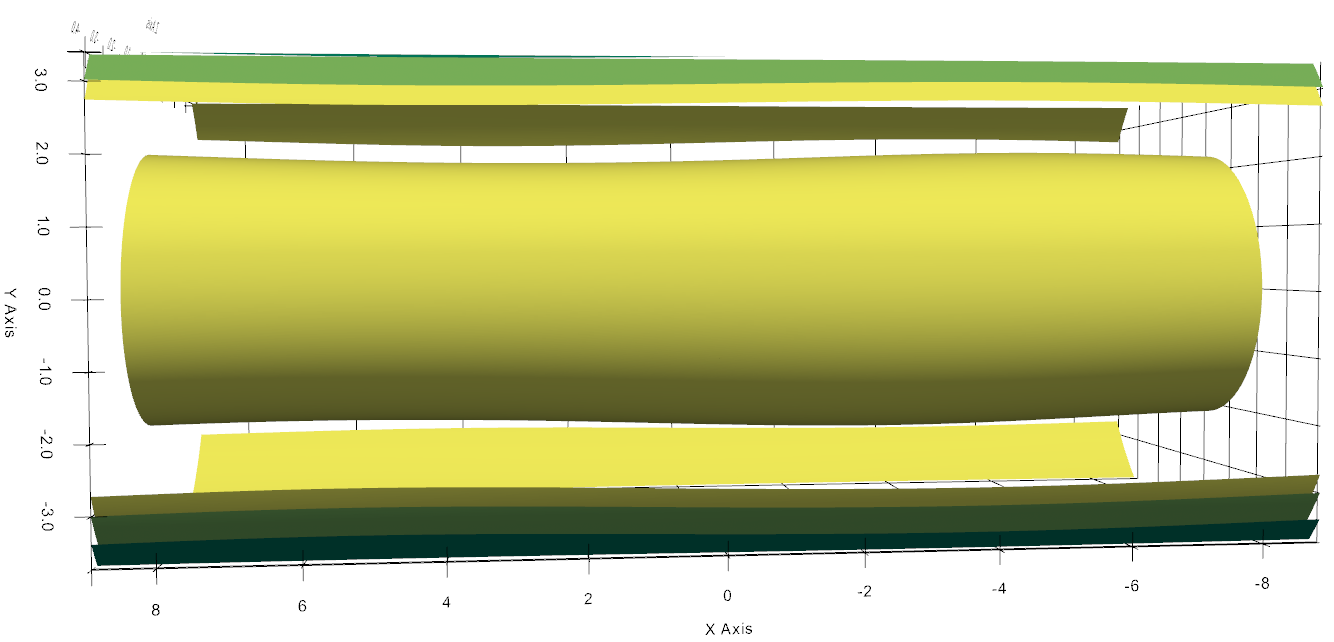}
        \caption{Initial condition, $t=0$.}\label{subfig:mode0_t0}
    \end{subfigure}
    \begin{subfigure}{0.32\textwidth}
        \centering
        \includegraphics[width=\textwidth]{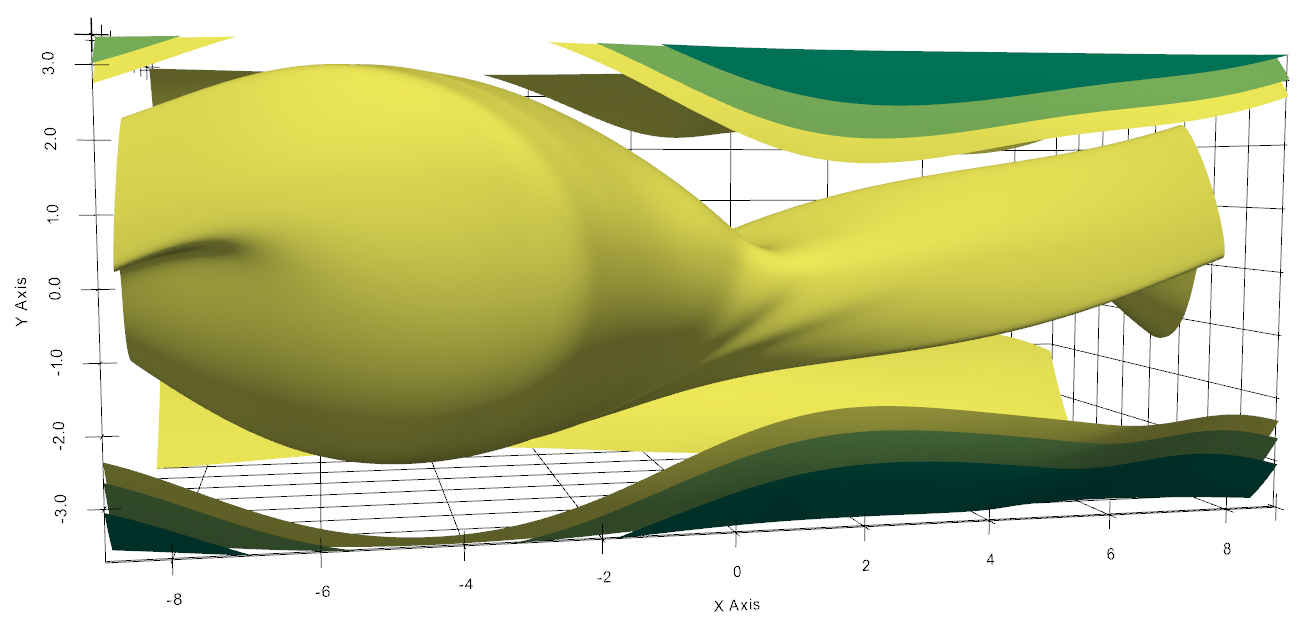}
        \caption{Developed trapping, $t=80$.}\label{subfig:mode0_t1}
    \end{subfigure}
    \begin{subfigure}{0.32\textwidth}
        \centering
        \includegraphics[width=\textwidth]{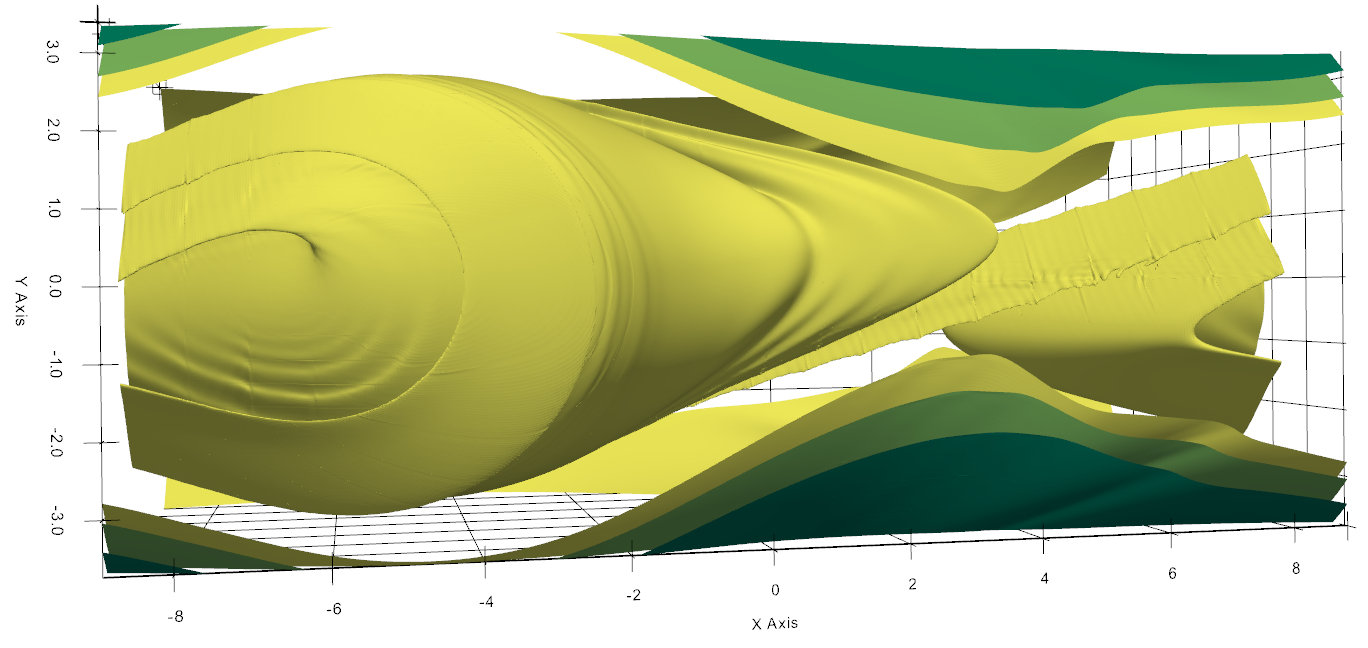}
        \caption{Further filamentation, $t=120$.}\label{subfig:mode0_t2}
    \end{subfigure}
    \caption{Phase space view in $(X,Y,Z)\equiv (x,u,v)$ of simulation $A$ focused on $(x,u)$-plane as
    iso-contours at $15\%$ of $\text{max}(f)$ (yellow).
    Only the inner domain is shown with $u,v\in (-4, 4)$, so the outer edges are visible.
    A trapped particle structure develops consisting of a ribbon of density winding around a
    a separatrix, while the outer ring consists of untrapped (passing) particles.}\label{fig:simulation_a_evolution}
\end{figure}

\begin{figure}
    \centering
    \begin{subfigure}{0.32\textwidth}
        \centering
        \includegraphics[width=\textwidth]{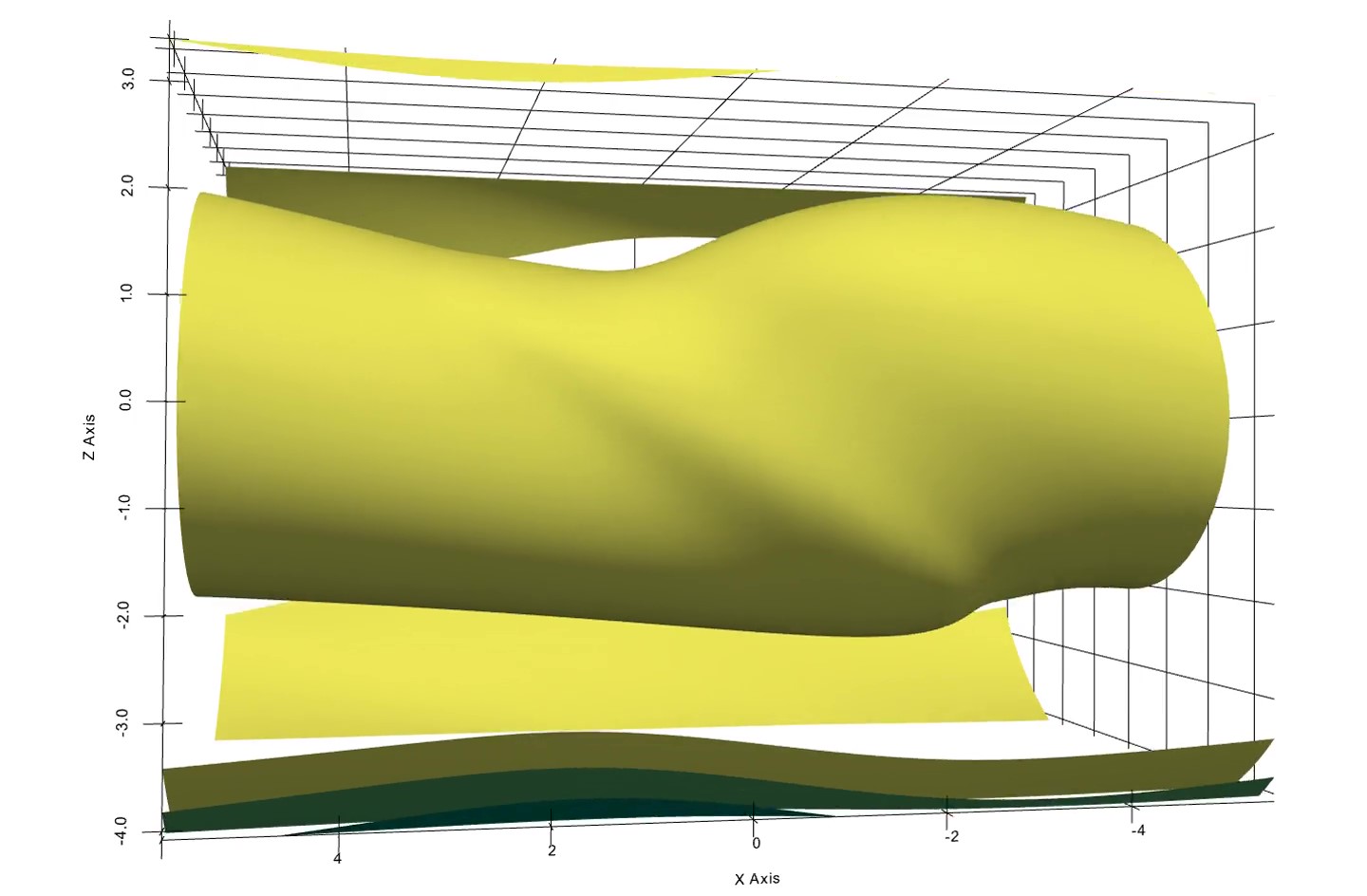}
        \caption{Developing mode, $t=100$.}\label{subfig:mode1_t0}
    \end{subfigure}
    \begin{subfigure}{0.32\textwidth}
        \centering
        \includegraphics[width=\textwidth]{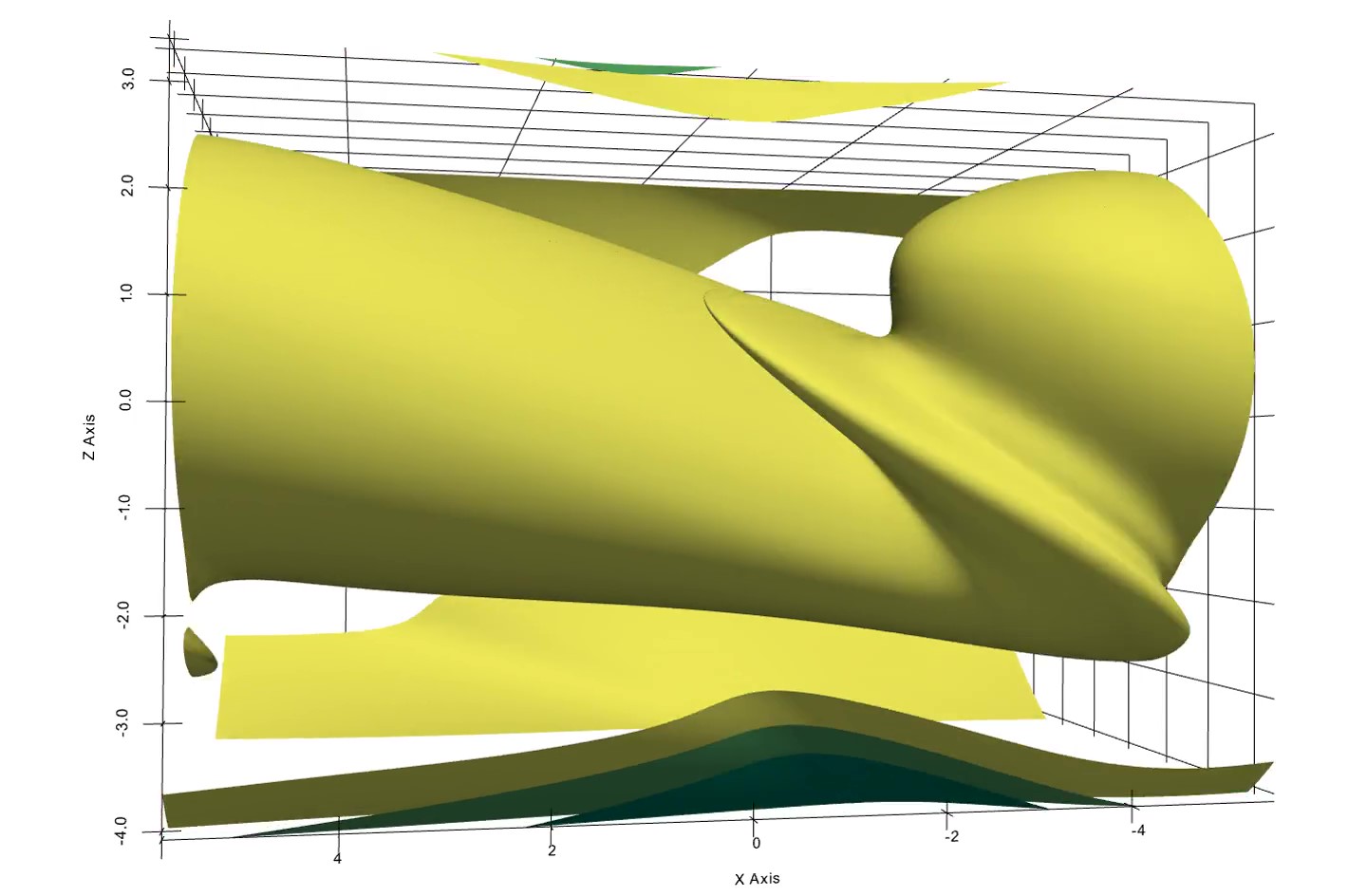}
        \caption{Developed vortex, $t=160$.}\label{subfig:mode1_t1}
    \end{subfigure}
    \begin{subfigure}{0.32\textwidth}
        \centering
        \includegraphics[width=\textwidth]{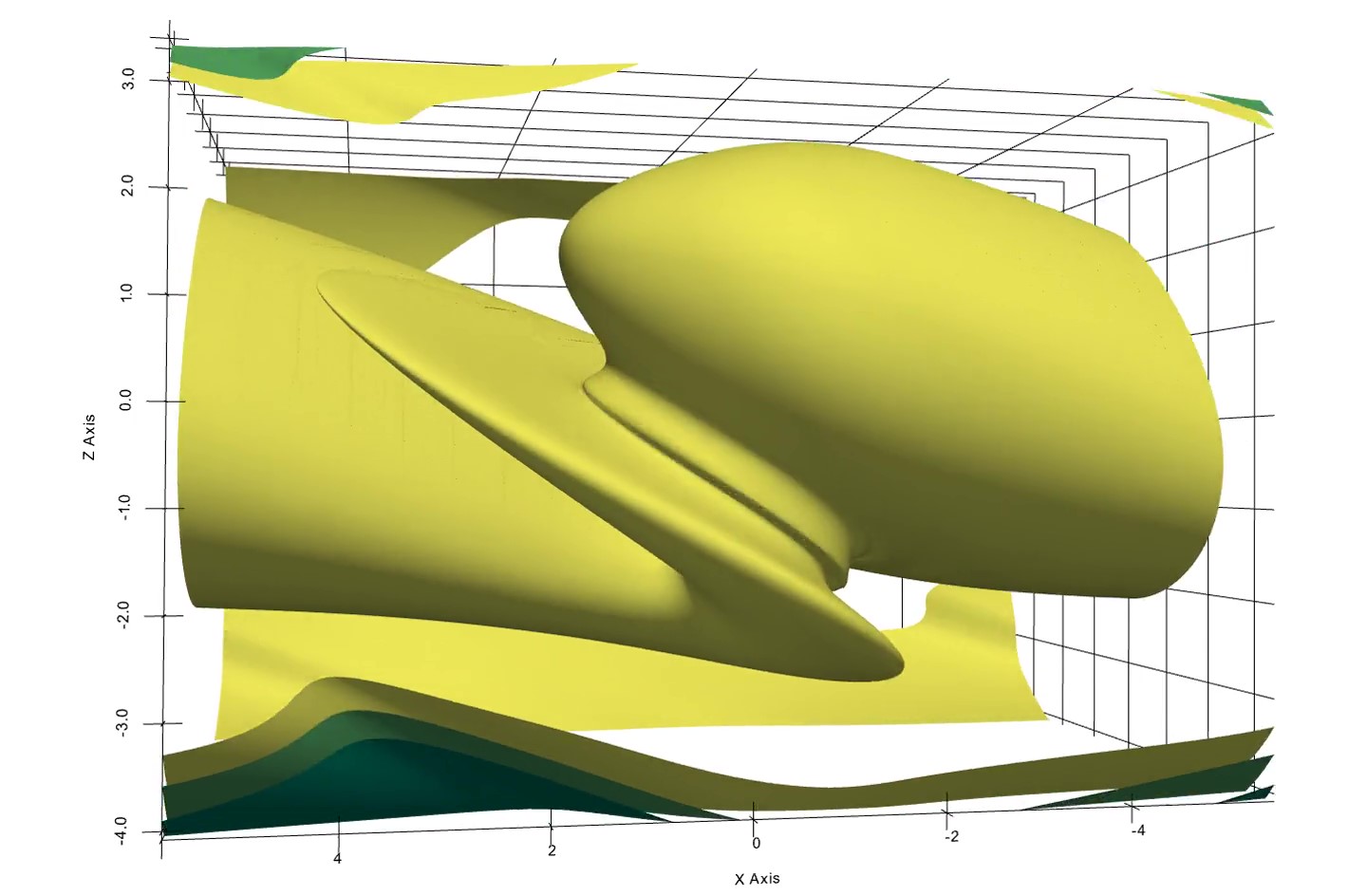}
        \caption{Translating vortex, $t=180$.}\label{subfig:mode1_t2}
    \end{subfigure}
    \caption{Phase space view looking on $(-x,v)$ - plane of simulation $B$ at $15\%$ iso-contours of $\text{max}(f)$
        (yellow). The mode is seen to be a growing, translating potential $\Phi(x)$ of positive phase velocity $v_{\varphi} = \omega_r / k$
        with an underlying phase space vortex structure centered at $(u,v)=0$.
        The vortex shape is explained by considering the trajectory of a test particle in the wave.
        That is, particles with a velocity close to that of the wave see a stationary potential and are accelerated to a
        high $u$-velocity.
        They then translate towards positive $x$ while their velocity vector is rotated by the Lorentz force
        to $-u$ at a rate close to the wave frequency (as $\omega_r \approx 1.2\omega_c$).
        The particle then advects opposite the direction of wave propagation, before repeating the cycle.}
    \label{fig:simulation_b_evolution}
\end{figure}

Three-dimensional isosurface plots were produced using PyVista, a Python package for VTK. To prepare the data,
an average was first taken for nodes lying on element boundaries for smoothness, and the
$8$-nodes per element were resampled to $25$ linearly spaced points per axis and
per element on the basis functions of Section~\ref{sec:lobatto}.
These iso-contours are shown for simulations $A$ and $B$ in Figs.~\ref{fig:simulation_a_evolution} and~\ref{fig:simulation_b_evolution}
respectively.
Both cases result in phase space structures with fine features, a phenomenon in self-consistent kinetic dynamics called
filamentation~\cite{cheng}.
These filaments develop shortly into the saturated state, showing the importance of high-resolution
and high-order techniques in Eulerian simulation of Vlasov-Poisson systems.

\begin{figure}
    \centering
    \begin{subfigure}{0.45\textwidth}
        \centering
        \includegraphics[width=\textwidth]{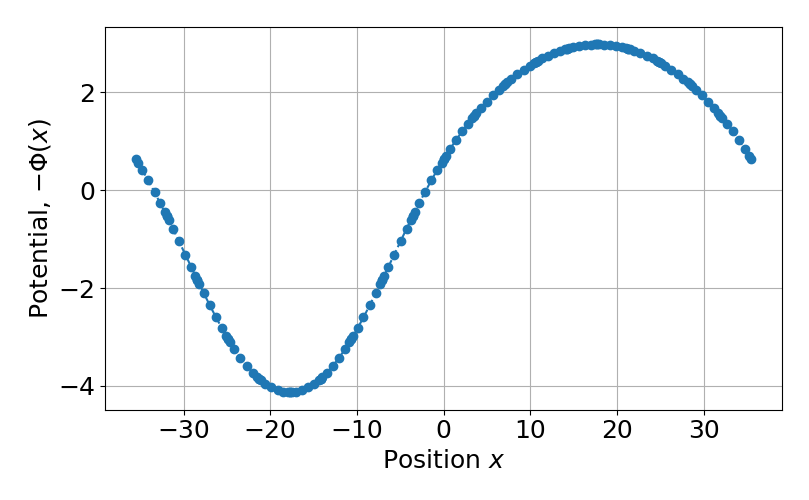}
        \caption{Simulation A at $t=120$.}\label{subfig:energyA}
    \end{subfigure}
    \begin{subfigure}{0.45\textwidth}
        \centering
        \includegraphics[width=\textwidth]{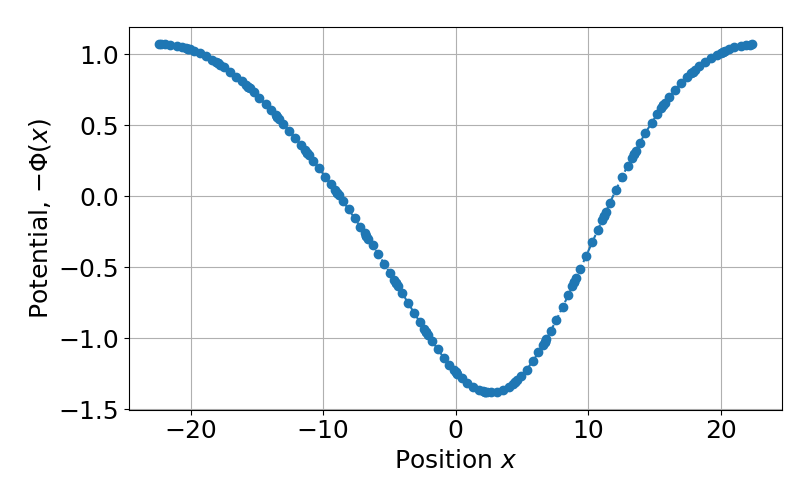}
        \caption{Simulation B at $t=180$}\label{subfig:potentialB}
    \end{subfigure}
    \caption{Electric potentials $\Phi(x)$ at saturation of the two studied cases. The potential of $A$
    is stationary while that of $B$ is translating to the right. The negative of the potential $-\Phi(x)$ is shown in order
        to account for the electron's negative charge. In both cases particle trapping structures develop in
    the potential wells, or minimum regions, in $-\Phi(x)$.}\label{fig:potentials}
\end{figure}

Figure~\ref{fig:potentials} shows the electric potentials $\Phi(x)$ in simulations $A$ and $B$.
In simulation $A$ the wave potential $\Phi(x)$ is stationary with a weakly fluctuating boundary,
so that part of the density $f(x,v)$ within the potential well executes trapped orbits.
This results in a trapping structure with orbits tracing a nonlinear potential similar to that seen in
electrostatic two-stream instability simulations where the potential is like that of a pendulum,
\textit{i.e.} $\Phi(x) = \sin(x)$, with its characteristic separatrix structure.
In this case particles also execute cyclotron motion so that the separatrix of the potential in simulation $A$ is
similar in form to that shown by the isosurfaces.

On the other hand, the saturated wave potential $\Phi(x)$ of simulation $B$ translates with positive phase velocity
$v_{\varphi} \approx \omega_r / k$.
The region of particle interaction translates along with the wave potential and forms a vortex structure with
the appearance of a kink in the phase space density $f(x,v)$.
The center of this kink continues to tighten as the simulation progresses, leading to progressively
finer structures just as in simulation $A$.
This effect is in agreement with the filamentation phenomenon and introduces simulation error as the structures
lead to large gradients on the grid scale where discreteness produces dispersion error.
For this reason the simulation is stopped at $t = 200$.

\begin{figure}
    \centering
    \begin{subfigure}{0.49\textwidth}
        \centering
        \includegraphics[width=\textwidth]{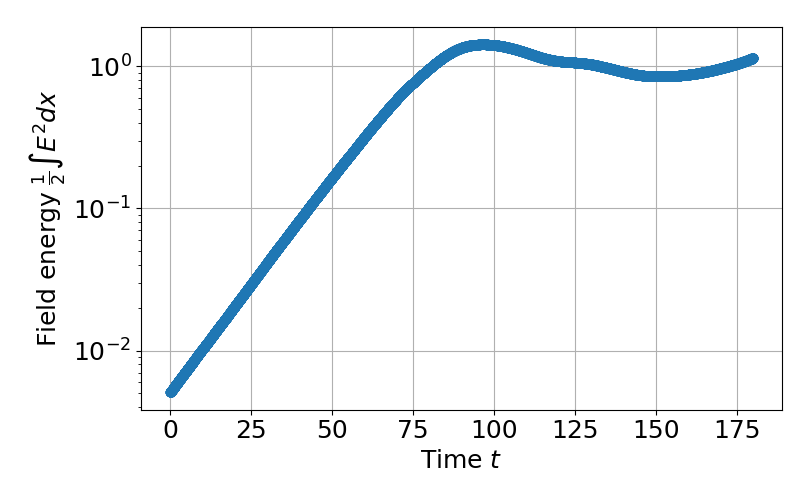}
        \caption{Simulation A, with $\omega_A = 0.349i$.}\label{subfig:energyA}
    \end{subfigure}
    \hfill
    \begin{subfigure}{0.49\textwidth}
        \includegraphics[width=\textwidth]{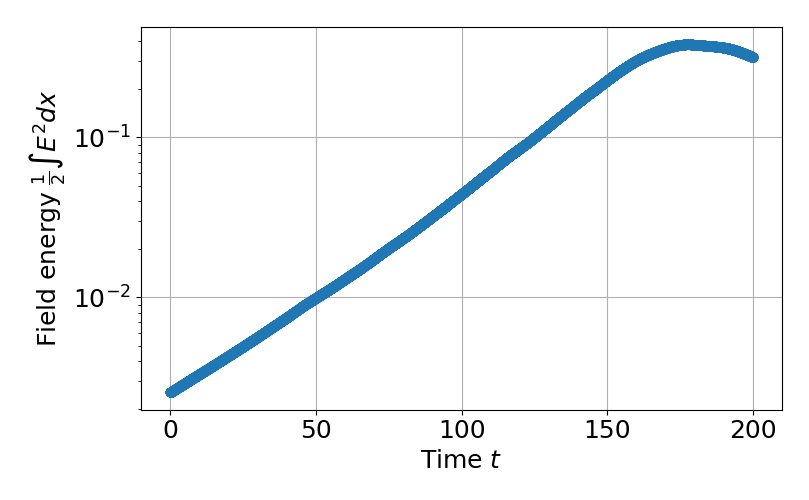}
        \caption{Simulation B, with $\omega_B = 1.18 + 0.131i$}\label{subfig:energyB}
    \end{subfigure}
    \caption{Domain-integrated field energy growth profiles for simulations $A$ and $B$. The thermal energy of the
    zero-order distribution is $0.25$ per unit length, according to Appendix~\ref{sec:therm_sec} with $\alpha = 1$.
    In simulation $A$ this corresponds to a domain-integrated thermal energy of $E_A\approx 17.8$ and in simulation $B$
    to $E_B\approx 11.2$. In both cases the instability saturates with an electric energy a few percent of the
        thermal energy, approximately $6\%$ in $A$ and $2.5\%$ in $B$.}\label{fig:energy}
\end{figure}

\section{Conclusion}\label{sec:conclusion}
This study is a self-contained description of how to implement the discontinuous Galerkin method on the GPU,
\textit{i.e.} a technique to solve PDEs using many finite elements, where the approximation in each element is supported
by interpolation polynomials.
When these basis polynomials interpolate the Gauss-Lobatto quadrature points of the one-dimensional
interval $\mathcal{I}\equiv[-1,1]$ then explicit and interpretable forms of the discontinuous Galerkin operators
can be found.
This is the case due to the affinity of quadrature on $\mathcal{I}$ to the spherical harmonics through the
theory of Legendre polynomials.
Recognizing this affinity allows the calculation of connection coefficients to alternative basis sets;
for example in this work the Fourier spectrum of the interpolation polynomials is utilized to solve the Poisson equation
to $\mathcal{O}((\Delta x)^{n+1/2})$ accuracy.

An important finding is that orthogonal discretization of a computational domain
results in significant computational savings over simplex elements, and increasingly so per dimension.
That is, only one-dimensional basis matrices need be constructed for general orthogonal discretizations
using $n$-cube elements built up from tensor products of $\mathcal{I}$, while full tensors must be constructed
for simplex elements.
Finally, these results are combined to solve an example problem,
namely the Vlasov-Poisson system of coupled hyperbolic-elliptic PDEs, using
GPU-accelerated libraries.
The semi-discrete equation is evaluated in only a few lines of code by utilizing CUDA wrappers with NumPy-like data arrays,
allowing tensor-product index ordering in a simple routine.
This approach does not outperform a custom implementation with CUDA code~\cite{einkemmer}, yet it has the advantage
for beginners of simplicity.



\bibliography{main}
\bibliographystyle{unsrt}\label{null:document}

\appendix
    \section{Implemention of upwind fluxes in Python with broadcasting}\label{sec:upwind_appendix}
    This appendix details how the upwind numerical flux may be prepared for the basis product function
    described in Section~\ref{subsec:semi_discrete_with_python}.
    The function \texttt{cupy.roll(a, shift)} (viz. \texttt{numpy.roll}) returns a view into a shifted axis of
the tensor array, and is ideal to compute the numerical flux using GPU-accelerated element-wise operations.
In each direction $j$ the numerical flux array $\mathcal{F}^{\alpha,\beta,j}$ is to be constructed of shape
\texttt{[$N_0, n_0, N_1, n_1, \cdots, N_j, 2, \cdots, N_{d-1}, n_{d-1}$]} for $N_j$ the elements in $j$ and $n_d$
its sub-element nodes.
The advection speed array, of size \texttt{[$N_j, n_j$]}, has its sign measured by the two arrays
\begin{verbatim}
    one_negatives = cp.where(condition=speed < 0, x=1, y=0)
    one_positives = cp.where(condition=speed > 0, x=1, y=0)
\end{verbatim}
which calculates $a^R$ and $a^L$.
The numerical flux array for each direction can then be built with broadcasting.
To do this in a dimension-independent fashion, define \texttt{boundary\_slices} for element boundaries and
\texttt{advection\_slices} for the flux vector components.

In two dimensions, for example, \texttt{boundary\_slices} is defined as the list of tuples
\begin{verbatim}
    e0, e1 = slice(elements[0]), slice(elements[1])
    n0, n1 = slice(nodes[0]), slice(nodes[1])
    self.boundary_slices = [
            # x-directed face slices [(left), (right)]
            [(e0, 0, e1, n1), (e0, -1, e1, n1)],
            # y-directed face slices [(left), (right)]
            [(e0, n0, e1, 0), (e0, n0, e1, -1)] ]
\end{verbatim}
On the other hand, \texttt{advection\_slices} is an array to prepare broadcasting with one of two directions of the
\textit{e.g.} shape \texttt{[elements[0], nodes[0], elements[1], 2]} numerical flux array $\mathcal{F}^{\alpha,\beta,j}$ for direction 1.
For example, if this 2D flux were for a rotation $F=\begin{bmatrix} -y, x\end{bmatrix}^T$, the advection slices are
\begin{verbatim}
    self.advection_slices = [(None, slice(elements[1]), slice(nodes[1])),
                             (slice(elements[0]), slice(nodes[0]), None)]
\end{verbatim}
Having set up this infrastructure, the numerical flux in direction \texttt{dim} is constructed by the function:
\begin{verbatim}
    def flux_in_dim(self, flux, one_negatives, one_positives, basis_matrix, dim):
        # allocate
        num_flux = cp.zeros(self.num_flux_sizes[dim])

        # Upwind flux, left face (with surface normal vector -1.0)
        num_flux[self.boundary_slices[dim][0]] =
            -1.0 * (cp.multiply(cp.roll(flux[self.boundary_slices[dim][1]],
                                        shift=1, axis=self.grid_axis[dim]),
                                one_positives[self.advection_slices[dim]]) +
                    cp.multiply(flux[self.boundary_slices[dim][0]],
                                one_negatives[self.advection_slices[dim]]))

        # Upwind flux, right face (with surface normal vector +1.0)
        num_flux[self.boundary_slices[dim][1]] =
            (cp.multiply(flux[self.boundary_slices[dim][1]],
                         one_positives[self.advection_slices[dim]]) +
             cp.multiply(cp.roll(flux[self.boundary_slices[dim][0]], shift=-1,
                                 axis=self.grid_axis[dim]),
                         one_negatives[self.advection_slices[dim]]))

        return basis_product(flux=num_flux, basis_matrix=basis.xi,
                             axis=self.sub_element_axis[dim],
                             permutation=self.permutations[dim])
\end{verbatim}
In the case of a hyperbolic problem, if the sign of the advection speed is constant during a problem then the sign
arrays should be computed prior to the main loop.

\section{Thermal properties of the ring distribution}\label{sec:therm_sec}
    This appendix discusses the first and second moments of the ring distribution
    \begin{equation}
        f_0(v) = \frac{1}{2\pi\alpha^2 j!}\Big(\frac{v^2}{2\alpha^2}\Big)^j\exp\Big(-\frac{v^2}{2\alpha^2}\Big)
    \end{equation}
    where $v$ is the radial part in polar coordinates.
    The first radial moment $\langle v\rangle$ measures the average velocity of the particles in the ring.
The moment is given by
    \begin{align}
        \langle v\rangle &\equiv \int_0^{\infty} v f_0(v) (2\pi v)dv\\
        &=\frac{2}{j!}\int_0^{\infty}\Big(\frac{v^2}{2\alpha^2}\Big)^{j+1}\exp\Big(-\frac{v^2}{2\alpha^2}\Big)dv\\
        &= \sqrt{2}\alpha\frac{\Gamma(j+\frac{3}{2})}{\Gamma(j+1)}
    \end{align}
    by using Euler's integral, where $\Gamma(j+1) = j!$ is the Gamma or factorial function.
    Expanding the Gamma ratio in Laurent series about $j=\infty$ yields
    \begin{equation}
        \frac{\Gamma(j+\frac{3}{2})}{\Gamma(j+1)} = \sqrt{j}\Big(1 + \frac{3}{8j} + \mathcal{O}(j^{-2})\Big) \approx \sqrt{j}
    \end{equation}
    so that $\langle v\rangle \approx \sqrt{2j}\alpha$, which corresponds with the peak of the distribution.
    Likewise, the centered second moment $\frac{1}{2}\langle (v - \langle v\rangle)^2\rangle$ measures the average energy of the
    particles. It is given by
    \begin{align}
        \frac{1}{2}\langle (v - \langle v\rangle)^2\rangle &= \frac{1}{2}\int_0^{\infty} (v - \langle v\rangle)^2 f_0(v)(2\pi v)dv,\\
                                                &= \frac{1}{2}\Big(\Big[\int_0^{\infty} v^2 f_0(v)(2\pi v)dv\Big] -
        2\langle v\rangle^2 + \langle v\rangle^2\Big),\\
                                                &= \alpha^2(j + 1 - \Big(\frac{\Gamma(j+\frac{3}{2})}{\Gamma(j+1)}\Big)^2).
    \end{align}
    Expanding this result about $j=\infty$ in Laurent series, one has
    \begin{equation}
        j + 1 - \Big(\frac{\Gamma(j+\frac{3}{2})}{\Gamma(j+1)}\Big)^2 = \frac{1}{4}\Big(1 - \frac{1}{8j} +
        \mathcal{O}(j^{-2})\Big) \approx \frac{1}{4}.
    \end{equation}
    The average velocity and thermal energy of the ring distribution can be taken to a fair degree of accuracy
as (with leading-order accuracy of $3/(8j)$ and $1/(8j)$ respectively),
    \begin{align}
        \langle v\rangle &\approx \sqrt{2j}\alpha,\\
        \frac{1}{2}\langle(v - \langle v\rangle)^2\rangle &\approx \Big(\frac{\alpha}{2}\Big)^2.
    \end{align}
\end{document}